\documentclass[english,prd,superscriptaddress,longbibliography,nofootinbib,preprint]{revtex4-2}
\usepackage[T1]{fontenc}
\usepackage[utf8]{inputenc}
\usepackage{color}
\usepackage{babel}
\usepackage{mathrsfs}
\usepackage{amsmath}
\usepackage{amssymb}
\usepackage{graphicx}
\usepackage[pdfusetitle,
 bookmarks=true,bookmarksnumbered=false,bookmarksopen=false,
 breaklinks=false,pdfborder={0 0 1},backref=false,colorlinks=false]
 {hyperref}

\makeatletter

\usepackage{mathrsfs}
\usepackage{braket}
\usepackage{bbm}

\makeatother

\begin{document}
\title{Many-body atomic response functions of xenon and germanium for leading-order
sub-GeV dark matter-electron interactions in effective field theory}
\author{C.-P.~Liu}
\email{cpliu@mail.ndhu.edu.tw}

\affiliation{Department of Physics, National Dong Hwa University, Shoufeng, Hualien
97401, Taiwan}
\author{Mukesh~K.~Pandey}
\email{mkpandey@gmail.com}

\affiliation{Department of Physics, Center for Theoretical Physics, and Leung Center
for Cosmology and Particle Astrophysics, National Taiwan University,
Taipei 10617, Taiwan}
\affiliation{Department of Physics, National Dong Hwa University, Shoufeng, Hualien
97401, Taiwan}
\author{Lakhwinder~Singh}
\affiliation{Department of Physics, Central University of South Bihar, Gaya 824236,
India}
\affiliation{Institute of Physics, Academia Sinica, Taipei 11529, Taiwan}
\author{Chih-Pan~Wu}
\email{jpw750811@gmail.com}

\affiliation{Department of Physics, National Dong Hwa University, Shoufeng, Hualien
97401, Taiwan}
\affiliation{Département de physique, Université de Montréal, Montréal H3C 3J7,
Canada}
\author{Jiunn-Wei~Chen}
\email{jwc@phys.ntu.edu.tw}

\affiliation{Department of Physics, Center for Theoretical Physics, and Leung Center
for Cosmology and Particle Astrophysics, National Taiwan University,
Taipei 10617, Taiwan}
\affiliation{Physics Division, National Center for Theoretical Sciences, National
Taiwan University, Taipei 10617, Taiwan}
\affiliation{Center for Gravitational Physics and Quantum Information, Yukawa Institute
for Theoretical Physics, Kyoto University, Kyoto 606-8502, Japan}
\author{Hsin-Chang~Chi}
\affiliation{Department of Physics, National Dong Hwa University, Shoufeng, Hualien
97401, Taiwan}
\author{Henry~T.~Wong}
\affiliation{Institute of Physics, Academia Sinica, Taipei 11529, Taiwan}
\begin{abstract}
Direct searches of dark matter candidates with mass energies less
than 1 GeV is an active research field. The energy depositions are
comparable to the scale of atomic, molecular, or condensed matter
systems, therefore many-body physics plays an important role in understanding
the detector's response in dark matter scattering. We present in this
work a comprehensive data set of atomic response functions for xenon
and germanium with 12.2 and 80 eV energy thresholds, respectively,
using the (multiconfiguration) relativistic random phase approximation.
This approach takes into account the relativistic, exchange, and correlation
effects in one self-consistent framework, and is benchmarked successfully
by photoabsorption data from thresholds to 30 keV with $\lesssim5\%$
errors. Comparisons with our previous and some other independent particle
approaches in literature are made. It is also found that the spin-dependent
(SD) response has significant difference from the spin-independent
(SI) one such that the dark matter SD and SI interactions with electrons
can be distinguished in unpolarized scattering, which is typical for
direct search detectors. Finally, the exclusion limits set by current
experiments are updated with our new results. 
\end{abstract}
\maketitle

\section{Introduction}

Dark matter (DM) particles with masses smaller than GeV have been
a subject receiving growing attention in recent years. They are well-motivated
in theory, and their detection methods are within the reach of current
and possible future experiments~\citep{Battaglieri:2017aum,Essig:2022dfa}.
Because of their small kinetic energies, $\lesssim O(\textrm{keV})$,
direct detectors must have sub-keV energy thresholds, for which electron
recoil (ER) is easier to be recorded than nuclear recoil. Among various
mechanisms that can trigger ER signals, DM-impact ionization through
DM-electron interactions is one of the most exploited search modes,
and stringent exclusion limits have been set~\citep{Kopp:2009et,Essig:2012yx,Roberts:2016xfw,Essig:2017kqs,Agnes:2018oej,Crisler:2018gci,Agnese:2018col,Pandey:2018esq,Abramoff:2019dfb,Roberts:2019chv,XENON:2019gfn,Aguilar-Arevalo:2019wdi,Catena:2019gfa,Arnaud:2020svb,Barak:2020fql,Amaral:2020ryn,PandaX-II:2021nsg,Liu:2021avx,XENON:2021qze,CDEX:2022kcd,DAMIC-M:2023gxo,SENSEI:2023zdf,Liang:2024lkk,Liang:2024ecw}

A key theory input for making an exclusion limit is the detector response
functions, which lead to a prediction of the differential count rate
at a detector with a prescribed DM interaction. As these low energy
scales overlap with the ones of atomic, molecular, or even condensed-matter,
many-body problems become an inevitable challenging task in constructing
the relevant response functions. In this paper, we only focus on the
kinematic regimes where detector responses can be well-described by
the single-atom pictures.

There have been quite a few atomic calculations in literature that
yield the response functions due to DM-electron interactions~\citep{Kopp:2009et,Essig:2011nj,Essig:2017kqs,Pandey:2018esq,Roberts:2019chv,Catena:2019gfa,Liu:2021avx,Hamaide:2021hlp,Caddell:2023zsw,Liang:2024lkk,Liang:2024ecw,Ge:2025itf}. However, the discrepancies
show that the predicted DM differential count rates depend sensitively on the
underlying many-body approaches. While all works share a common ground at the
mean field level, they differ on the formulations of the potentials used to
compute the final scattered state: an atomic ion plus one free electron.
Therefore, benchmarking an atomic many-body approach with some known processes
is an important justification.

In our previous studies, we have successfully applied an \textit{ab
initio} method: (multi-configuration) relativistic random phase approximation~\citep{Johnson:1979wr,Huang:1981wj},
(MC)RRPA, to photoionization of atomic (germanium) xenon and achieved
excellent agreement with photoabsorption data. However, applying the
same method to DM-impact ionization is both numerically challenging
and intensive. Therefore, in our previous papers~\citep{Pandey:2018esq,Liu:2021avx},
we resorted to a version of relativistic frozen core approximation
(RFCA) for the final states, which does not include the exchange potential,
nor electron-electron correlation beyond mean field. For selected
energy transfers, we did perform (MC)RRPA calculations, and reported
a general agreement at $20\%$ except at low energies.

In this work, we overcome the numerical challenges. Consequently the
resulting (MC)RRPA response functions of atomic xenon (germanium)
would cover the full parameter space for direct sub-GeV DM searches
with an energy threshold of 12.2 (80) eV. These atomic response functions
incorporate not only the important ingredients being identified previously:
relativistic correction~\citep{Roberts:2015lga,Roberts:2016xfw,Pandey:2018esq,Roberts:2019chv,Liu:2021avx}
and exchange potential~\citep{Hamaide:2021hlp,Caddell:2023zsw},
but furthermore, electron-electron correlation which is missing in
previous treatments based on the independent particle picture. As
will be shown in this paper, the correlation effect can be significant
when scattering energy transfer is below 100 eV.

It is important to note that to analyze experiments using very-low-threshold germanium
semiconductor detectors, one further needs condensed-matter many-body methods, for example,
(most popularly) the density functional theory, in order to exhibit the band structures
of the valence and conduction electrons and understand their response to DM scattering with
a energy deposition $\lesssim$ 100~$\textrm{eV}$ (see, e.g., Refs. ~\citep{Essig:2015cda,Knapen:2021bwg,Griffin:2021znd}).
On the other hand, when energy
deposition becomes large enough so that core electrons also get involved, their contributions
can be efficiently calculated by atomic methods, as these localized electrons are atomic-like.
Our setting of 80 eV as the applicable energy threshold for germanium detectors is justified
by benchmarking with photoabsroption data.

The paper is organized as follows. In Sec.~\ref{sec:response}, we
present the main results of this work: the atomic response functions
of xenon and germanium by (MC)RRPA. Using photoabsorption data as
benchmarks, we compare the (MC)RRPA and RFCA results, and show the
substantial improvement of the former as a justification of (MC)RRPA.
In Sec.~\ref{sec:cs=00003D000026rate}, we gather the essential formulae
that are needed to compute the differential count rates based on the
response function tables published along with the paper, and the codes
that automatize the processes are also supplied. In Sec.~\ref{sec:comp=00003D000026dis},
we compare and discuss the differences of the new (MC)RRPA results
for DM-impact ionization from our previous RFCA and a few others.
The exclusion plots of DM-electron interactions are updated with current
experiment data sets, and we summarize the paper in Sec.~\ref{sec:summary}.

\section{Response Functions~}\label{sec:response}

\subsection{Definition}

Light dark matter, for its small kinetic energy, can trigger an observable
electron recoil (ER) signal easier than a nuclear recoil (NR) signal,
therefore, the detector response to the DM-electron ($\chi$-$e$)
interaction is our primary concern. From the conventional effective-field-theory
construction~\citep{Kopp:2009et,Fan:2010gt,Fitzpatrick:2012ix,DelNobile:2013sia,Anand:2013yka,Anand:2014kea,Chen:2015pha,DelNobile:2018dfg,Catena:2019hzw,Trickle:2020oki},
the leading-order (LO) interaction Lagrangian 
\begin{align}
\mathscr{L}_{\chi\textrm{-}e}^{(\textrm{LO})}= & (c_{1}+d_{1}/q^{2})\left(\chi^{\dagger}\mathbbm{1}_{\chi}\chi\right)\cdot\left(e^{\dagger}\mathbbm{1}_{e}e\right)\nonumber \\
 & +(c_{4}+d_{4}/q^{2})\left(\chi^{\dagger}\vec{S}_{\chi}\chi\right)\cdot\left(e^{\dagger}\vec{S}_{e}e\right)\,,\label{eq:L_EFT}
\end{align}
contains a spin-independent (SI) and a spin-dependent (SD) parts,
where $\mathbbm{1}_{e(\chi)}$ and $\vec{S}_{e(\chi)}$ are the unity
and spin operators for the electron (DM) field, respectively. The
constants $c_{1}$ and $c_{4}$ denote the strengths for the short-ranged
interaction terms, while $d_{1}$ and $d_{4}$ for the long-ranged
counterparts where $q$ is the magnitude of the 3-momentum transfer.

For a complex atom with $Z$ electrons, the possible atomic transitions
are governed by four transition operators in coordinate space: 
\[
\mathbbm{1}_{e}\rightarrow\sum_{i=1}^{Z}e^{i\vec{q}\cdot\vec{r}_{i}}\mathbbm{1}_{i}\,,\vec{S}_{e}\rightarrow\sum_{i=1}^{Z}e^{i\vec{q}\cdot\vec{r}_{i}}{\frac{\vec{\sigma}_{i}}{2}}\,.
\]
The $2\times2$ unity and Pauli matrices, $\mathbbm{1}_{i}$ and $\vec{\sigma}_{i}$,
act on the nonrelativistic wave function of the $i$th electron. To
take into account the relativistic corrections, they are replaced
by $\mathbbm{1}_{i}^{\textrm{D}}=\left(\begin{array}{cc}
\mathbbm{1}_{i} & 0\\
0 & \mathbbm{1}_{i}
\end{array}\right)$ and $\vec{\sigma}_{i}^{\textrm{D}}=\left(\begin{array}{cc}
\vec{\sigma}_{i} & 0\\
0 & \vec{\sigma}_{i}
\end{array}\right)$ when atomic wave functions are in a Dirac 4-spinor form.

While Cartesian operators look compact in form, for actual atomic
many-body calculations, it is advantageous to transform them into
spherical multipoles. They are

\begin{subequations}
\begin{align}
\hat{{C}}_{J}^{M_{J}}(q) & =\sum_{i=1}^{Z}j_{J}(qr_{i})Y_{J}^{M_{J}}(\Omega_{r_{i}})\mathbbm{1}_{i}^{\textrm{D}}\,,\\
\hat{\Sigma}_{J}^{M_{J}}(q) & =\sum_{i=1}^{Z}j_{J}(qr_{i})\vec{Y}_{JJ}^{M_{J}}(\Omega_{r_{i}})\cdot\vec{\sigma}_{i}^{\textrm{D}}\,,\\
\hat{\Sigma}_{J}^{'M_{J}}(q) & =\sum_{i=1}^{Z}\left\{ -\sqrt{\frac{J}{2J+1}}j_{J+1}(qr_{i})\vec{Y}_{JJ+1}^{M_{J}}(\Omega_{r_{i}})\right.\nonumber \\
 & \left.+\sqrt{\frac{J+1}{2J+1}}j_{J-1}(qr_{i})\vec{Y}_{JJ-1}^{M_{J}}(\Omega_{r_{i}})\right\} \cdot\vec{\sigma}_{i}^{\textrm{D}}\,,\\
\hat{\Sigma}_{J}^{''M_{J}}(q) & =\sum_{i=1}^{Z}\left\{ \sqrt{\frac{J+1}{2J+1}}j_{J+1}(qr_{i})\vec{Y}_{JJ+1}^{M_{J}}(\Omega_{r_{i}})\right.\nonumber \\
 & \left.+\sqrt{\frac{J}{2J+1}}j_{J-1}(qr_{i})\vec{Y}_{JJ-1}^{M_{J}}(\Omega_{r_{i}})\right\} \cdot\vec{\sigma}_{i}^{\textrm{D}}\,,
\end{align}
\end{subequations}
 where $j_{J}(qr)$ is the spherical Bessel function, $Y_{J}^{M_{J}}(\Omega_{r})$
the spherical harmonics of solid angle $\Omega_{r}$, and $\vec{Y}_{JL}^{M_{J}}(\Omega_{r})$
the vector spherical harmonics formed by recoupling of $Y_{L}^{M_{L}}(\Omega_{r})$
and the unit vector $\hat{r}$, whose spherical projection is proportional
to $Y_{1}^{\lambda}$.~\footnote{For these multipole operators and their corresponding response functions,
we use the same convention as Refs.~\citep{Donnelly:1979ezn,Fitzpatrick:2012ix}.}

When the atomic initial state is unpolarized and the final polarization
state is not detected (i.e., summed), the algebra on the total magnetic
quantum numbers can be greatly simplified. In combination with total
angular momentum and parity selection rules, there is no interference
from the scattering amplitudes of these four multipole operators.
Therefore, we define four distinct atomic response functions 
\begin{align}
\mathscr{R}_{O_{J}}(T,q)= & \sum_{FJ_{F}}\overline{\sum_{IJ_{I}}}\left|\left\langle F,J_{F}\left\Vert \hat{O}_{J}(q)\right\Vert I,J_{I}\right\rangle \right|^{2}\nonumber \\
 & \times\delta(E_{\mathscr{F}}-E_{\mathscr{I}}-T)\,,\label{eq:response_def}
\end{align}
with $\hat{O}_{J}$ being one of $(\hat{{C}}_{J},\hat{\Sigma}_{J},\hat{\Sigma}_{J}^{'},\hat{\Sigma}_{J}^{''})$.
The atomic initial (final) state $\ket{\mathscr{I}(\mathscr{F})}$
is completely specified by its total angular momentum and $z$-projection,
$J_{I(F)}$ and $M_{J_{I(F)}}$, and other quantum numbers collectively
labeled by $I(F)$. The double-bared, or reduced, matrix element notation
and the missing of $M_{J,J_{I(F)}}$ indicate the reduction of angular
momentum algebra. The bar over the initial state sum means the average
of the ground state configurations. The delta function imposes the
energy conservation: $T$ is the energy deposition by the DM particle
which causes the ionization of a bound electron, plus a tiny atomic
center-of-mass recoil: $q^{2}/(2m_{A})$ where $m_{A}$ is the atomic
mass. For a non-relativistic incident DM particle, the atomic recoil
is only important at high energy transfers which also require high
momentum transfers. After the sum and average is completed, one clearly
sees the response function only depends on two variables: the energy
and momentum transfer by DM.

\subsection{(MC)RRPA and RFCA Results}

In this work, the atomic initial, i.e., ground, state $\ket{I,J_{I}M_{J_{I}}}$
is obtained by solving the Dirac-Fock (DF) equation for closed-shell
atoms like xenon. For open-shell atoms like germanium, an additional
feature, the multiconfiguration (MC) of the reference state, is implemented
that yields the MCDF equation. The atomic final, i.e., ionized, state
$\ket{F,J_{F}M_{J_{F}}}$, is obtained by solving the corresponding
relativistic random phase approximation (RRPA) equation for xenon,
and the MCRRPA equation for germanium, respectively. Compared with
our previous works~\citep{Pandey:2018esq,Liu:2021avx} that used
RFCA~\footnote{In Refs.~\citep{Pandey:2018esq,Liu:2021avx}, we used the acronym
FCA for the method. Here we add an adjective ``relativistic'' to
better characterize it and make distinction from the corresponding
nonrelativistic version.} for the final state, the (MC)RRPA approach,~\footnote{The abbreviation (MC)RRPA refers to RRPA for closed-shell and MCRRPA
for open-shell atoms, respectively.} while being much more time-consuming, is not only self-consistent
with the exchange term built in, but also includes electron-electron
correlation beyond mean field through RPA. Details of (MC)RRPA can
be found in, e.g., Refs.~\citep{Johnson:1979wr,Huang:1981wj}. In
Appendix, we give an outline of how the wave functions and transition
amplitudes are computed in the (MC)RRPA scheme.

\subsubsection{Photoabsorption benchmark~}\label{subsec:photoabs}

Many-body correlation, which is beyond typical mean-field type approaches,
has been known to play an important role in proper understanding of
excited states of a many-body system. For an atom system, its photoabsorption
cross section, which is dominated by photoelectron emission in the
energy range of 10 eV to 100 keV, provides an ideal testing ground.

In radiation gauge, the photoabsorption process goes
through two types of vector current multipole operators, transverse
electric, $\hat{E}_{J}^{M_{J}}$, and transverse magnetic $\hat{M}_{J}^{M_{J}}$:
\begin{subequations}
\begin{align}
\hat{E}{}_{J}^{M_{J}}(q) & =\sum_{i=1}^{Z}\frac{1}{q}\,\overrightarrow{\nabla}_{i}\times[j_{J}(qr_{i})\vec{Y}_{JJ}^{M_{J}}(\Omega_{r_{i}})]\cdot\vec{\varsigma}_{i}^{\textrm{D}}\,,\\
\hat{M}_{J}^{M_{J}}(q) & =\sum_{i=1}^{Z}j_{J}(qr_{i})\vec{Y}_{JJ}^{M_{J}}(\Omega_{r_{i}})]\cdot\vec{\varsigma}_{i}^{\textrm{D}}\,,
\end{align}
\end{subequations}
where $\vec{\varsigma}_{i}^{\textrm{D}}=\left(\begin{array}{cc}
0 & \vec{\sigma}_{i}\\
\vec{\sigma}_{i} & 0
\end{array}\right)$. In terms of their corresponding response functions $\mathscr{R}_{E_{J}}$
and $\mathscr{R}_{M_{J}}$, the total cross section of absorbing a
photon of energy $T$ is given by the relation 

\begin{align}
\sigma_{\mathrm{abs}}^{(\gamma)}(T) & =\dfrac{2\pi^{2}\alpha}{T}\,\mathscr{R}_{\perp}^{(\gamma)}(T,q=T)\,,\\
\mathscr{R}_{\perp}^{(\gamma)}(T,q) & =\dfrac{4\pi}{2J_{I}+1}\sum_{J\geq1}^{\infty}\left[\mathscr{R}_{E_{J}}(T,q)+\mathscr{R}_{M_{J}}(T,q)\right]\,.\label{eq:S_abs}
\end{align}
The symbol ``$\perp$'' of $\mathscr{R}_{\perp}$denotes the transverse
character of the associated response function, and because the photon
is real, only the on-shell part of the response function $q=T$ is
probed by photoabsorption.

\begin{figure*}
\begin{centering}
\begin{tabular}{cc}
\includegraphics[width=0.45\textwidth]{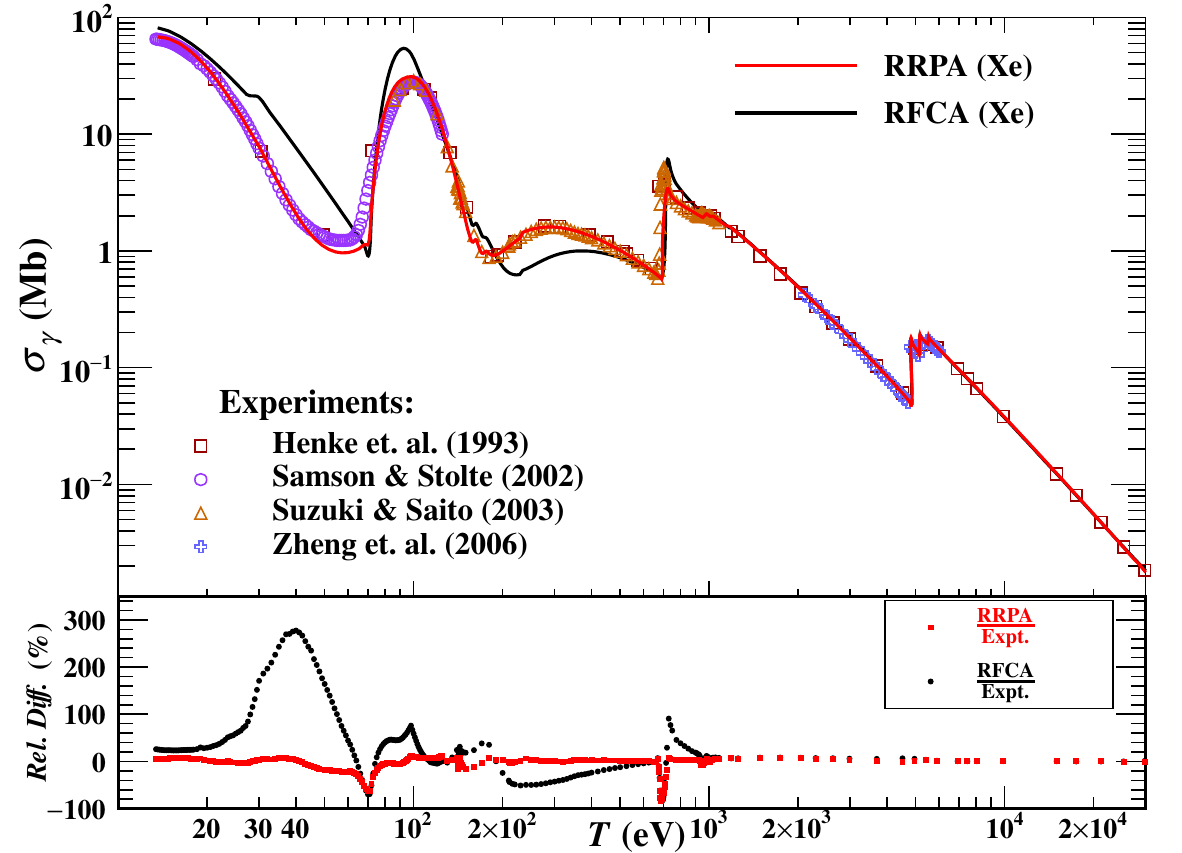}  & \includegraphics[width=0.45\textwidth]{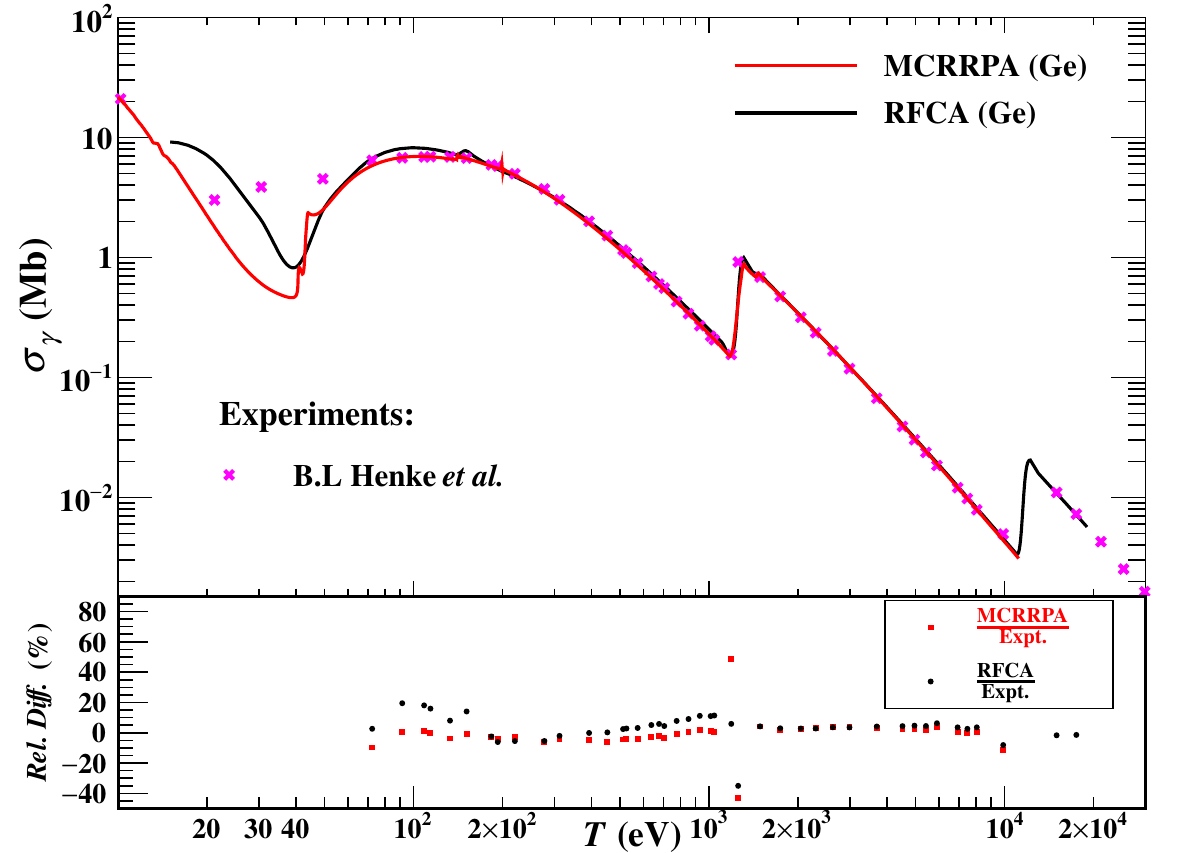} \tabularnewline
\end{tabular}
\par\end{centering}
\caption{The photoabsorption cross sections of xenon (left) and germanium (right)
are shown. The solid red and black lines indicate the results of our
(MC)RRPA and RFCA calculations, respectively. The germanium results
are compared with Ref.~\citep{HENKE1993181}, and the xenon results
are compared with Refs.~\citep{HENKE1993181,SAMSON2002265,SUZUKI200371,ZHENG2006143}.
The percentage differences of (MC)RRPA and RFCA calculations from
the experimental data are shown in bottom insets.~}\label{fig:comp_photo}
\end{figure*}

In the left panel of Fig.~\ref{fig:comp_photo}, we show the comparison
of our RRPA and RFCA results for xenon with experimental data compiled
from Refs.~\citep{HENKE1993181,SAMSON2002265,SUZUKI200371,ZHENG2006143}.
In our new RRPA run, we overcome several numerical difficulties in the
energy range of 70-90 eV. As a result, we improve on our previous
RRPA calculations reported in Ref.~\citep{Chen:2016eab} such that
the new RRPA calculations pass the benchmark in the entire energy
range from threshold up to 30 keV. Except when $T$ is near the photoabsorption
edges, i.e., ionization thresholds, the general agreement between
RRPA and experiments is within $5\%$. An even more important feature
shown in this figure is the comparison between the RRPA and RFCA results.
While RFCA agrees with experiments well for $T\gtrsim1\,\textrm{keV}$,
for lower energies, there are noticeable discrepancies.

The range between 70 eV to 1 keV was an important testing ground for
atomic calculations historically. It was Copper who first proposed
a qualitative solution using a simple central field potential for
the $4d$ electrons~\citep{Cooper:1964cm}, whose ionizations dominate
the cross section. While later approaches such as refined central
potential~\citep{Manson:1968pi} and Hartree-Fock approximation~\citep{Kennedy:1972pi}
yield better agreements, they failed to describe correctly the shapes
of the two peaks and the positions of two maxima (one at 100 and the
other at 291eV) and one minimum (193 eV), due to the missing of correlation
effect. Similarly for $T<70\,\textrm{eV}$, the correlation effect
is also important for the ionization of $5p$ and $5s$ orbitals and
was first demonstrated in Ref.~\citep{Johnson:1979pi} by applying
RRPA to a limited subshells including $5p$, $5s$, and $4d$.
With modern computing resources, the valence electron configuration
in our RRPA run includes all except two innermost $1s$ electrons,
whose high ionization energy, $\sim35\,\textrm{keV}$, justify their
inert character in low energy processes. This excellent benchmark
not only justifies the robustness of our approaches, but also indicate
the necessity of a genuinely many-body response function for xenon
detector at the energy range of sub-keV.

In the right panel of Fig.~\ref{fig:comp_photo}, the comparison
of our MCRRPA results for germanium (all electrons including $1s$
are treated as valence) with experimental data compiled from Ref.~\citep{HENKE1993181}
is essentially the same as reported in Ref.~\citep{Chen:2013lba}.
Except in $T\lesssim80\,\textrm{eV}$, where our single-atom calculation
misses the crystal band structures of $3d$, $4s$, and $4p$ orbitals,
the inner core states of germanium semiconductor are highly localized
and their dynamics in photoionization can be accurately described
by MCRRPA. As a result, we only provide response functions for germanium
for $T\ge80\,\textrm{eV}$. On the other hand, the discrepancy between
RFCA and MCRRPA shown in this figure for $T\lesssim40\,\textrm{eV}$
gives another indication that correlation is important for outer shell
electrons including $3d$, $4s$, and $4p$ for atomic germanium.

\subsubsection{Data tables and parameter space covered}

Four types of atomic response functions: charge ($C$), axial longitudinal
($L^{5}$), axial transverse electric ($E^{5}$), and axial transverse
magnetic ($M^{5}$), $\mathscr{R}_{C,L^{5},E^{5},M^{5}}(T,q)$, which
correspond to transition operators $\hat{M},$ $\hat{\Sigma}^{''}$,
$\hat{\Sigma}^{'}$, $\hat{\Sigma}$, respectively, are compiled in
this work. The digital data files are shipped along with the codes
that supplement this paper. An excerpt from the charge response function
of xenon is reproduced in Table~\ref{tab:RF}. The first and the
second column give the DM energy and momentum transfer, $T$ and $q$
in units of eV. The third column gives the RRPA value in units of
1/eV. The RFCA response functions are also provided for reference,
and their corresponding values are listed in the fourth column.

\begin{table}
\begin{center}

\begin{tabular*}{1\textwidth}{@{\extracolsep{\fill}}cccc}
$T$ (eV) & $q$ (eV) & $\mathscr{R}_{C}^{(\textrm{RRPA})}$(eV$^{-1}$) & $\mathscr{R}_{C}^{(\textrm{RFCA})}$(eV$^{-1}$)\tabularnewline
\hline 
\hline 
12.2 &  4713.228991  &  3.147887e-02  &  4.145701e-02  \tabularnewline
12.2 &  5577.051911  &  2.974885e-02  &  3.153066e-02  \tabularnewline
12.2 &  6440.874832  &  2.225311e-02  &  2.174820e-02  \tabularnewline
12.2 &  7304.697752  &  1.377670e-02  &  1.282298e-02  \tabularnewline
12.2 &  8168.520673  &  7.319998e-03  &  6.346181e-03  \tabularnewline
12.2 &  9032.343593  &  3.417749e-03  &  2.556397e-03  \tabularnewline
12.2 &  9896.166514  &  1.441683e-03  &  7.762880e-04  \tabularnewline
12.2 &  10759.989434 &  5.939323e-04  &  1.618018e-04  \tabularnewline
12.2 &  11623.812355 &  2.890757e-04  &  7.420893e-05  \tabularnewline
12.2 &  12487.635275 &  1.943690e-04  &  1.456781e-04  \tabularnewline
$\vdots$ & $\vdots$ & $\vdots$ & $\vdots$\tabularnewline
\end{tabular*}
\end{center}

\caption{Excerpt from the data files for the charge response function of xenon,
calculated by the RRPA (3rd column) and RFCA (4th column) methods.~}\label{tab:RF}

\end{table}

The minimal values of $T$ for xenon and germanium are fixed at $T_{\min}=12.2$
and $80\,\textrm{eV}$, respectively. The former is the first ionization
energy of xenon, and the latter is the lower limit that atomic calculation
can be safely applied to semiconductor germanium. The maximal value
of $T$ is fixed at $T_{\max}\approx5\,\textrm{keV}$, which is roughly
the largest kinetic energy of a one-GeV DM particle in our galaxy.

The main reason why a (MC)RRPA calculation is much more time-consuming
than typical independent-particle ones is the evaluation of the matrix
element of $\hat{O}_{J}(q)$. For each operator of a specific $q$
and $J$, a (MC)RRPA equation is set up and needs to be solved self-consistently.
As a result, the momentum grid and the multipole expansion have to
be fixed economically.

For a given $T$, the momentum grid is determined based on the following
considerations. By the kinematics of galactic cold DM, the minimal
and maximal momentum transfer for a given speed $v_{\chi}$ are simply
\[
q_{\min}^{\max}=m_{\chi}v_{\chi}(1\pm\sqrt{1-T/(T_{\chi})})\,,
\]
respectively, where $T_{\chi}=\frac{1}{2}m_{\chi}v_{\chi}^{2}$ is
the NR kinetic energy of the DM particle. Given that $v_{\max}$ is
the maximum DM speed without escaping our galaxy, a $m_{\chi}$ independent
absolute threshold value can be fixed by $q_{\textrm{th}}=T/v_{\max}$.
The maximum momentum transfer which can be handled reliably by our
current (MC)RRPA codes is $2.5\,\textrm{MeV}$. Because of the fast
convergence in radial integrals involving the spherical Bessel function
of increasing $q$, the momentum grid has a denser sampling at the
low $q$ region, and the high momentum tail beyond $2.5\,\textrm{MeV}$
is extrapolated to the end point $q_{\textrm{end}}=\sqrt{2m_{A}(T-T_{\min})}$.

The multipole expansion is truncated at $J_{\max}$
where its contribution is on the order of $10^{-4}$ compared to the
biggest one in the series. For the current coverage of $(T,q)$, we
found that $J_{\max}\leq6$ is sufficient. This is supported by our
RFCA calculations which can easily handle high multipoles.

\begin{widetext}
\begin{figure*}
\begin{centering}
\begin{tabular}{cc}
\includegraphics[width=0.45\textwidth]{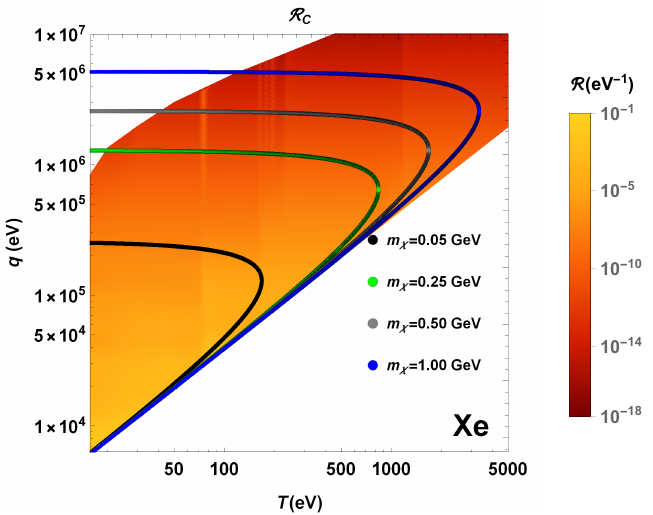}  & \includegraphics[width=0.45\textwidth]{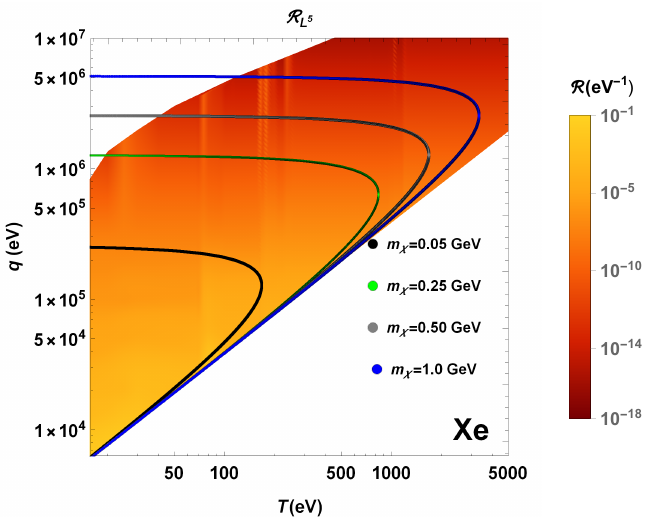}\tabularnewline
\includegraphics[width=0.45\textwidth]{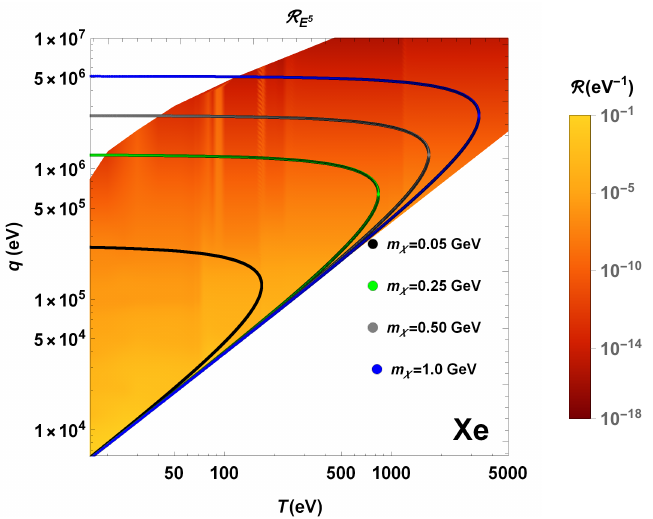}  & \includegraphics[width=0.45\textwidth]{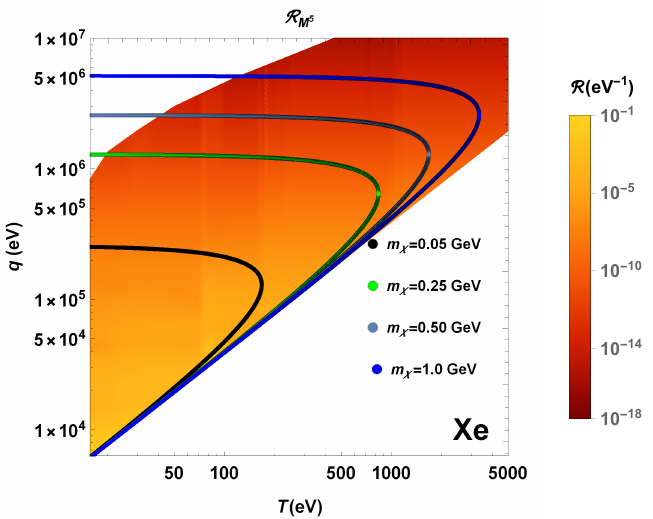}\tabularnewline
\end{tabular}
\par\end{centering}
\caption{Four atomic responses $\mathscr{R}_{C}$, $\mathscr{R}_{L^{5}}$,$\mathscr{R}_{E^{5}}$,
and $\mathscr{R}_{M^{5}}$ of xenon as functions of $T$ and $q$
calculated by RRPA with values shown by color gradients~}\label{fig:RF_Xe}
\end{figure*}
\end{widetext}

\begin{figure*}
\begin{centering}
\begin{tabular}{cc}
\includegraphics[width=0.45\textwidth]{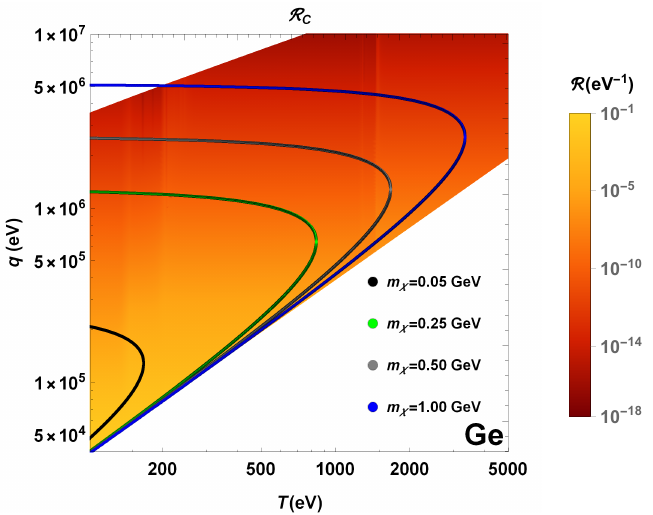}  & \includegraphics[width=0.45\textwidth]{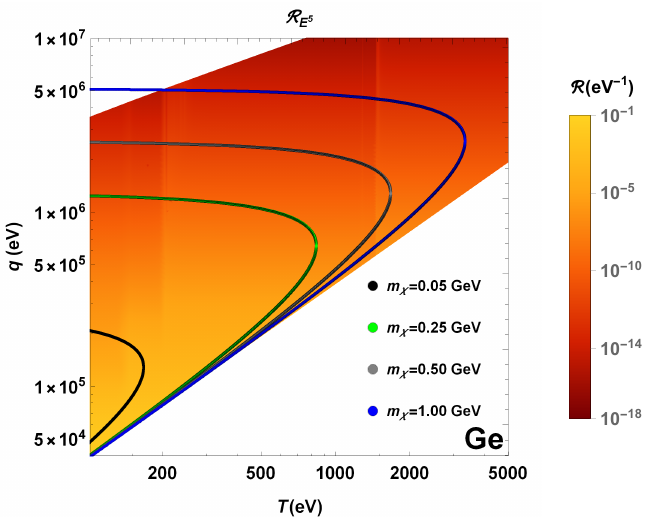}\tabularnewline
\includegraphics[width=0.45\textwidth]{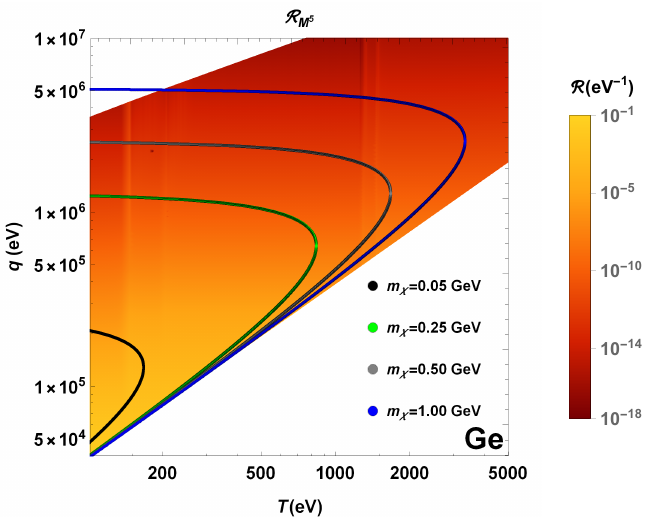}  & \includegraphics[width=0.45\textwidth]{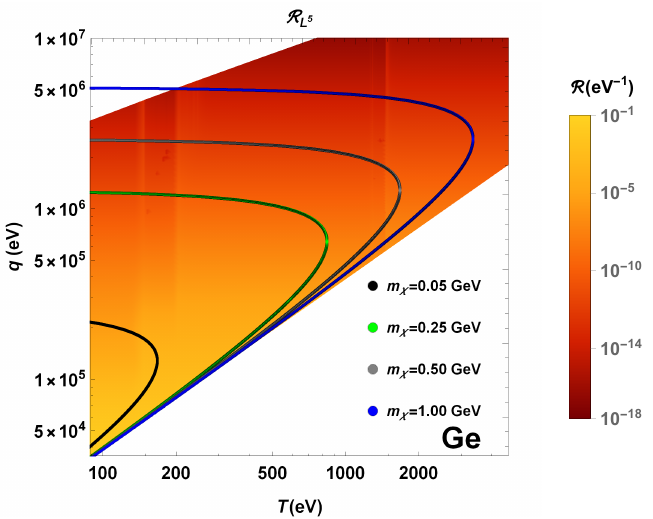}\tabularnewline
\end{tabular}
\par\end{centering}
\caption{Four atomic responses $\mathscr{R}_{C}$, $\mathscr{R}_{L^{5}}$,$\mathscr{R}_{E^{5}}$,
and $\mathscr{R}_{M^{5}}$ of germanium as functions of $T$ and $q$
calculated by MCRRPA with values shown by color gradients.~}\label{fig:RF_Ge}
\end{figure*}

In Figs.~\ref{fig:RF_Xe} and \ref{fig:RF_Ge}, the four response
functions $\mathscr{R}_{C}$, $\mathscr{R}_{L^{5}}$,$\mathscr{R}_{E^{5}}$,
and $\mathscr{R}_{M^{5}}$ for xenon and germanium, respectively,  are
plotted in two dimensional planes of$T$ and $q$ with color gradients
showing their magnitudes. There are two kinetically forbidden regions:
the lower right is to guarantee a large enough $q$ for an energy
transfer $T$, i.e., $q_{\textrm{th}}$ defined above. The upper left
is to cut off an atomic recoil $q^{2}/2m_{A}$ too large to leave
sufficient energy left for first ionization $E_{\textrm{ion}}$. The
four contours bound the allowed energy transfer, $T\in[0,\,1/2m_{\chi}v_{\max}^{2}]$,
and momentum transfer, $q\in[q_{\min},\,q_{\max}]$ depending on $T$
with $v_{\chi}=v_{\max}$, by scattering of four different dark matter
masses. As can be seen, the current data tables are sufficient to
cover dark matter searches with masses up to $\sim1\,\textrm{GeV}$,
and lower to $\sim10\,(80)\,\textrm{MeV}$ for xenon (germanium) detectors. 

\section{Differential Cross Sections and Rates~}\label{sec:cs=00003D000026rate}

\subsection{Formulation}

With the relevant response functions being obtained, it is straightforward
to assemble them and calculate differential cross sections and rates.
Here we outline the general procedure and essential formulae.

The differential cross section due to the LO $\chi$-$e$ interaction
for a DM particle of mass $m_{\chi}$ and speed $v_{\chi}$ is calculated
by an integration over momentum transfer $q$ 
\begin{align}
\frac{d\sigma}{dT}=\frac{1}{2\pi v_{\chi}^{2}} & \int_{q_{\min}}^{q_{\max}}\,qdq\,\Big\{(c_{1}+d_{1}/q^{2})^{2}\mathscr{R}_{\textrm{SI}}(T,q)\nonumber \\
 & +(\bar{c}_{4}+\bar{d}_{4}/q^{2})^{2}\mathscr{R}_{\textrm{SD}}(T,q)\Big\}\,,\label{eq:dcs}
\end{align}
where $(\bar{c}_{4},\bar{d}_{4})=\sqrt{s_{\chi}(s_{\chi}+1)/12}(c_{4},d_{4})$
with $s_{\chi}$ being the spin of the DM particle. The SI and SD
response functions are obtained by

\begin{subequations}
\begin{align}
\mathscr{R}_{\textrm{SI}}(T,q)=\frac{4\pi}{2J_{I}+1} & \sum_{J=0}\mathscr{R}_{C_{J}}(T,q)\,,\label{eq:response_SI}\\
\mathscr{R}_{\textrm{SD}}(T,q)=\frac{4\pi}{2J_{I}+1} & \left\{ \sum_{J=1}\left[\mathscr{R}_{E_{J}^{5}}(T,q)+\mathscr{R}_{M_{J}^{5}}(T,q)\right]\right.\nonumber \\
 & +\left.\sum_{J=0}\mathscr{R}_{L_{J}^{5}}(T,q)\right\} \,.\label{eq:response_SD}
\end{align}
\end{subequations}

From the differential cross section for a fixed DM speed $v_{\chi}$,
the differential count rate per single atom is computed by convoluting
with the DM number flux spectrum

\begin{equation}
\frac{dR}{dT}=n_{\chi}\int d^{3}v_{\chi}f(\vec{v}_{\chi})v_{\chi}\frac{d\sigma}{dT}\,,\label{eq:count-rate}
\end{equation}
where $n_{\chi}$ is the total galactic DM number density, and $f(\vec{v}_{\chi})$
is the three-dimensional DM velocity distribution with respect to
the Earth frame. The seasonal effect, which is due to the relative
Earth velocity to the local DM halo, $\vec{v}_{\textrm{E}}$, is built
in $f(\vec{v}_{\chi})$.

The advantages of using response functions can now be seen more clearly.
In the whole process of predicting a DM-atom ionization rate, the
inputs include (i) the DM spectrum from astrophysics and cosmology,
(ii) the DM-electron interaction from particle physics, and (iii)
high-quality wave functions from atomic physics. By packing up the
most computing-expensive part (iii) of the inputs in the form of response
functions, the studies of parts (i) and (ii) can be carried out using
the same atomic input.

\subsection{Accompanied Code and Test Examples~}\label{subsec:test_egs}

We supply computer codes, in both C and Mathematica, along with this
paper that read in the database of response functions and compute
the differential count rate in units of $\textrm{kg}^{-1}\textrm{keV}^{-1}\textrm{day}^{-1}$
through either the SI or SD $\chi$-$e$ interaction at leading order.~\footnote{The code and data tables can be downloaded through the link: \textcolor{red}{\url{https://web.phys.ntu.edu.tw/~jwc/DarkMatterandNeutrinoGroup/index.html?pg=AtomicResponses}.}}
For local DM velocity spectrum $f(\vec{v}_{\chi})$, we assume the
Maxwellian form of the standard halo model (see, e.g., Ref.~\citep{Freese:2012xd})
in this paper, with seasonal modulation averaged out. For all numerical
results, the local DM density $\rho_{\chi}$, circular speed $v_{0}$,
escape speed $v_{\textrm{esc}}$, and Earth speed $v_{E}$ are chosen
to be $0.4\,\textrm{GeV/cm}^{3}$, 220, 544, and 232 km/s, respectively.
For the test examples, we further fix $m_{\chi}=1\,\textrm{GeV}$,
and the $\chi$-$e$ coupling constants for the SR and LR interactions
to be $c_{1}=\bar{c}_{4}=1\,\textrm{GeV}^{-2}$ and $d_{1}=\bar{d}_{4}=10^{-9}$,
respectively.

Note that the above computations involve a double integration, first
over $q$ and second over $v_{\chi}$ (when the seasonal effect is
averaged). We can easily compare them with a commonly-used procedure
which reverses the integration order and assume the only $v_{\chi}$
dependence of $d\sigma/dT$ is $1/v_{\chi}^{2}$. This results in
the so-called mean inverse speed function 
\begin{equation}
\eta(\tilde{v}_{\textrm{min}})=\int d^{3}v_{\chi}f(\vec{v}_{\chi})\frac{1}{v_{\chi}}\Theta(v_{\chi}-\tilde{v}_{\min})\,,\label{eq:eta_func}
\end{equation}
which only integrates the DM velocity spectrum above a minimal value
\[
\tilde{v}_{\min}=\frac{T}{q}+\frac{q}{2m_{\chi}}\,,
\]
that can afford a given energy and momentum transfer at $T$ and $q$.
By this ansatz, the differential rate can alternatively be computed
by a single integration 
\begin{align}
\frac{dR}{dT}^{(\eta)} & =\frac{n_{\chi}}{2\pi}\int_{q_{\min}}^{q_{\max}}\,qdq\,\eta(\tilde{v}_{\textrm{min}})\Big\{(c_{1}+d_{1}/q^{2})^{2}\mathscr{R}_{\textrm{SI}}(T,q)\nonumber \\
 & +(\bar{c}_{4}+\bar{d}_{4}/q^{2})^{2}\mathscr{R}_{\textrm{SD}}(T,q)\Big\}\,,\label{eq:dR/dT_eta}
\end{align}
and is also implemented in the code.

\begin{figure}
\begin{centering}
\begin{tabular}{cc}
\includegraphics[width=0.45\textwidth]{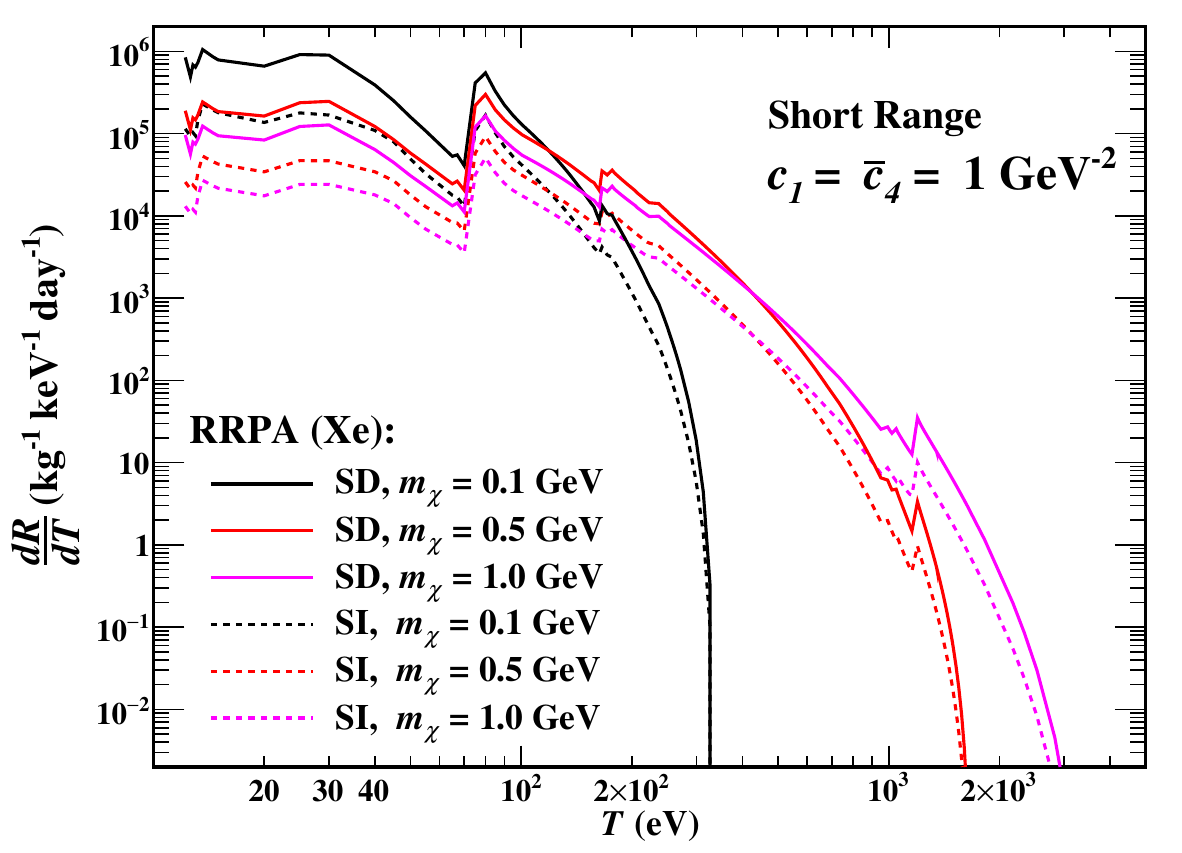}  & \includegraphics[width=0.45\textwidth]{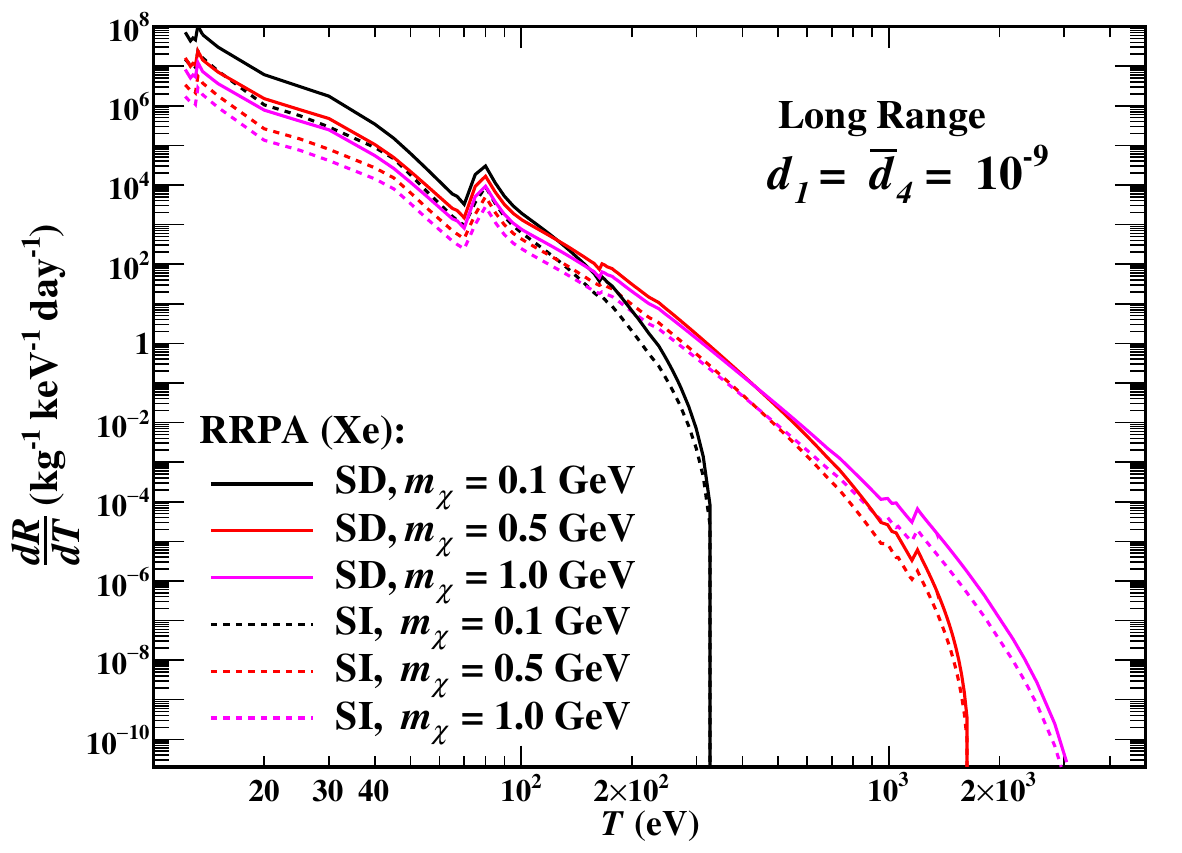}\tabularnewline
\end{tabular}
\par\end{centering}
\caption{Differential count rates $dR/dT$ due to the SI (dashed) or SD (solid)
$\chi$-$e$ short-range (left) or long-range (right) interaction
for a xenon detector with a kg-day exposure calculated by RRPA response
functions.~}\label{fig:test_examples}
\end{figure}

Test examples of predicted differential count rates for a xenon detector
with a kg-day exposure are summarized in Fig.~\ref{fig:test_examples}.
The agreement of $dR/dT$ and $dR^{(\eta)}/dT$ is found to be excellent.

\begin{figure*}
\begin{centering}
\begin{tabular}{cc}
\includegraphics[width=0.45\textwidth]{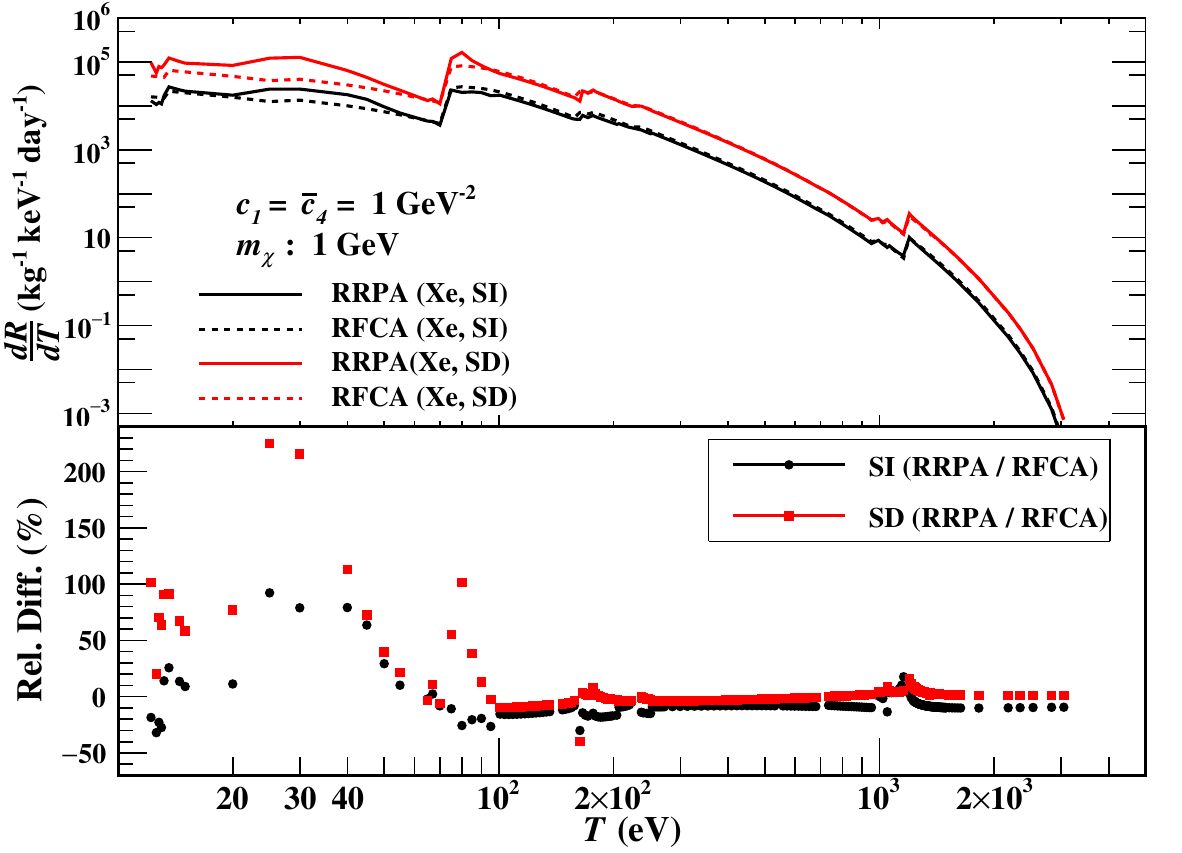}  & \includegraphics[width=0.45\textwidth]{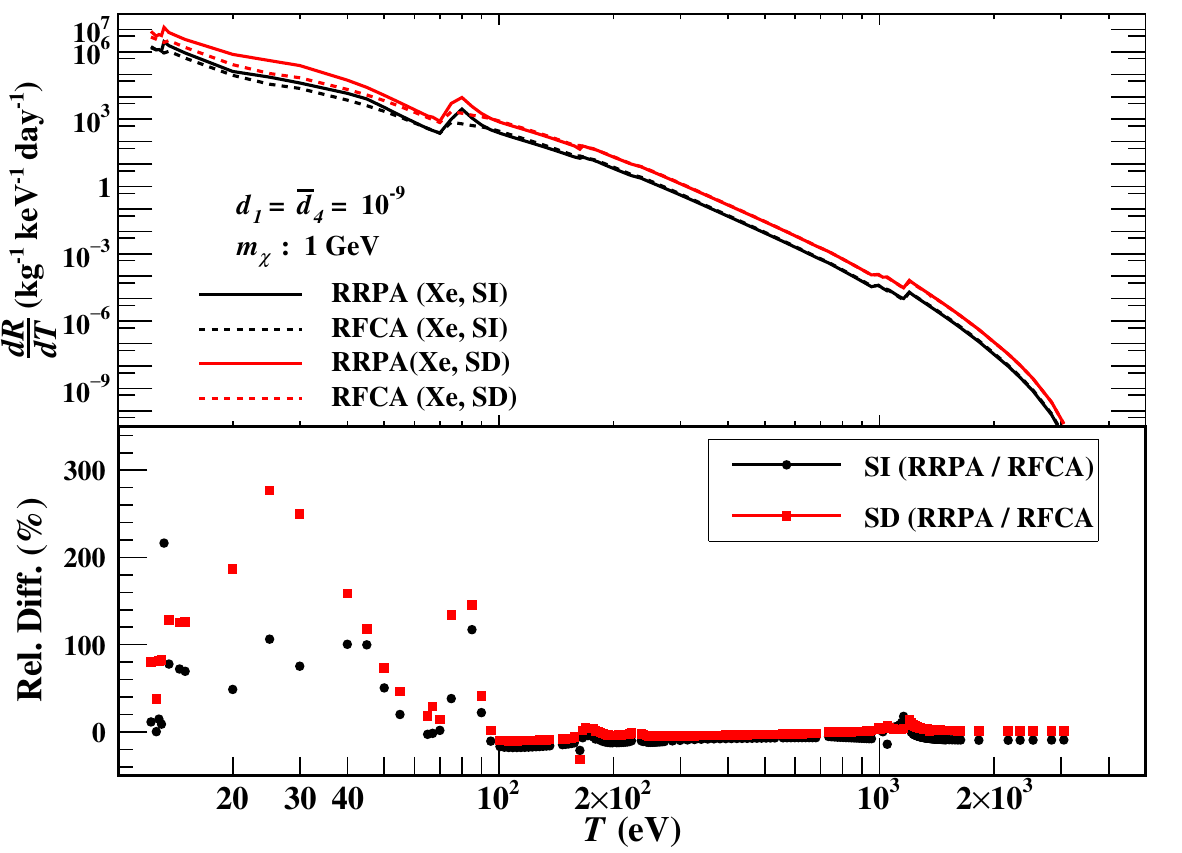}\tabularnewline
\end{tabular}
\par\end{centering}
\caption{Comparison of the RRPA with the RFCA predictions of $\frac{dR}{dT}$
for the $m_{\chi}=1\,\textrm{GeV}$ case in Fig.~\ref{fig:test_examples}.
The bottom inset shows their percentage difference defined in the
text.~\ref{subsec:test_egs}. }\label{fig:diff_RRPA/FCA}
\end{figure*}

\section{Comparisons, Discussions, and New Exclusion Limits~}\label{sec:comp=00003D000026dis}

The comparison of RRPA and RFCA photoionization cross sections with
experiments for xenon in Fig.~\ref{fig:comp_photo} has demonstrated
the combined effect from exchange and correlation. \textcolor{black}{(Taking
into comparison the works of Refs.~\citep{Manson:1968pi,Kennedy:1972pi},
which had exchange effect included, one can judge that the correlation
effect is more important.)} Now we further examine their differences
in the predictions of $dR/dT$ in DM-xenon (DM-germanium) scattering
in Fig.~\ref{fig:diff_RRPA/FCA} for the test examples given above.
In terms of percentage differences: 
\[
\left(\frac{dR/dT(\textrm{RRPA})}{dR/dT(\textrm{RFCA})}-1\right)\times100\%
\]
one can see that for $T\gtrsim300\,\textrm{eV}$, the agreement between
RRPA and RFCA is generally within $20\%$, except around the ionization
thresholds $\sim1\,\textrm{keV}$.

However, for $T\lesssim300\,\textrm{eV}$, one
clearly sees larger discrepancies. In this energy range, the electrons
of xenon are ionized from three outermost shells $5p$, $5s$, and
$4d$. From the previous discussion of how typical independent particle
treatments failed to achieve reasonable agreement with photoabsorption
data in the similar energy range, this is not unexpected. Furthermore,
because the exchange and correlation effects are both range-dependent,
their corrections to DM-impact ionization with the SR or LR DM-electron
interaction would differ, which can be seen by comparing the left
and right panels.

\begin{figure*}
\begin{centering}
\begin{tabular}{cc}
\includegraphics[width=0.45\textwidth]{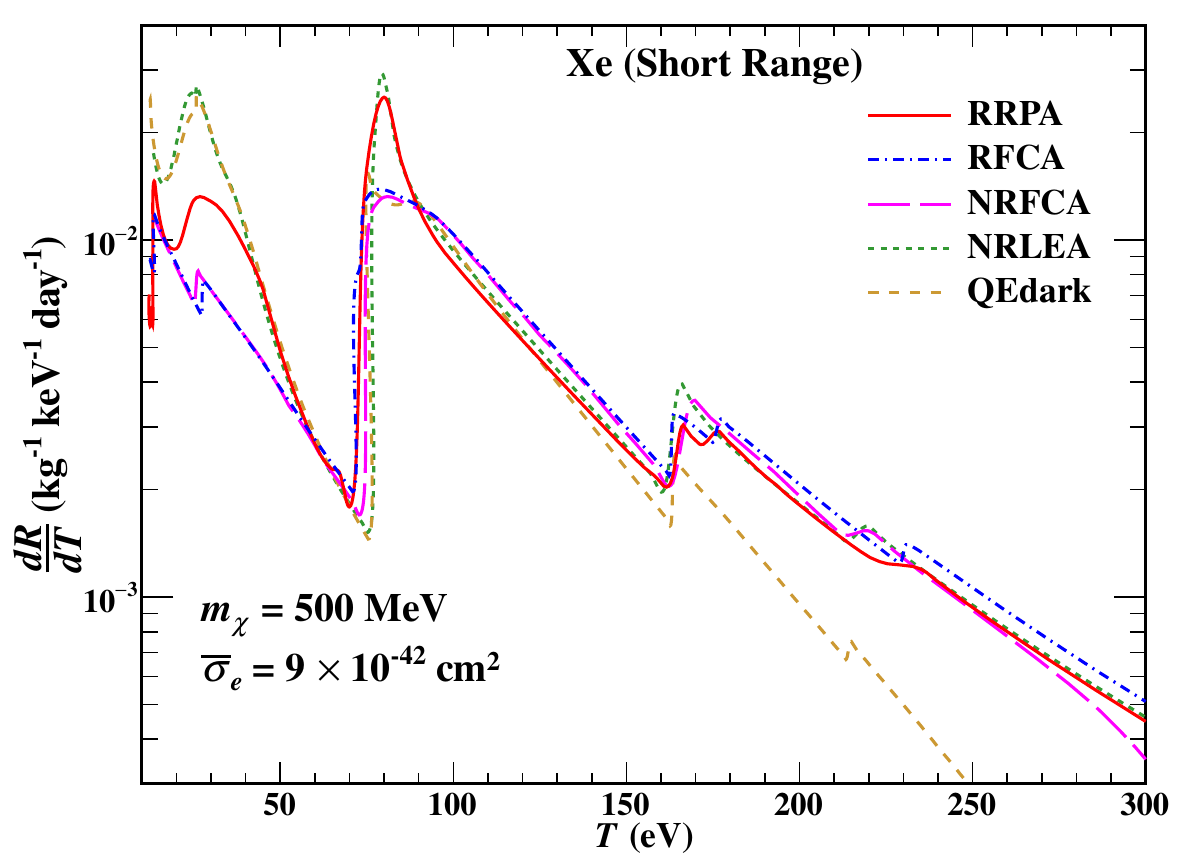}  & \includegraphics[width=0.45\textwidth]{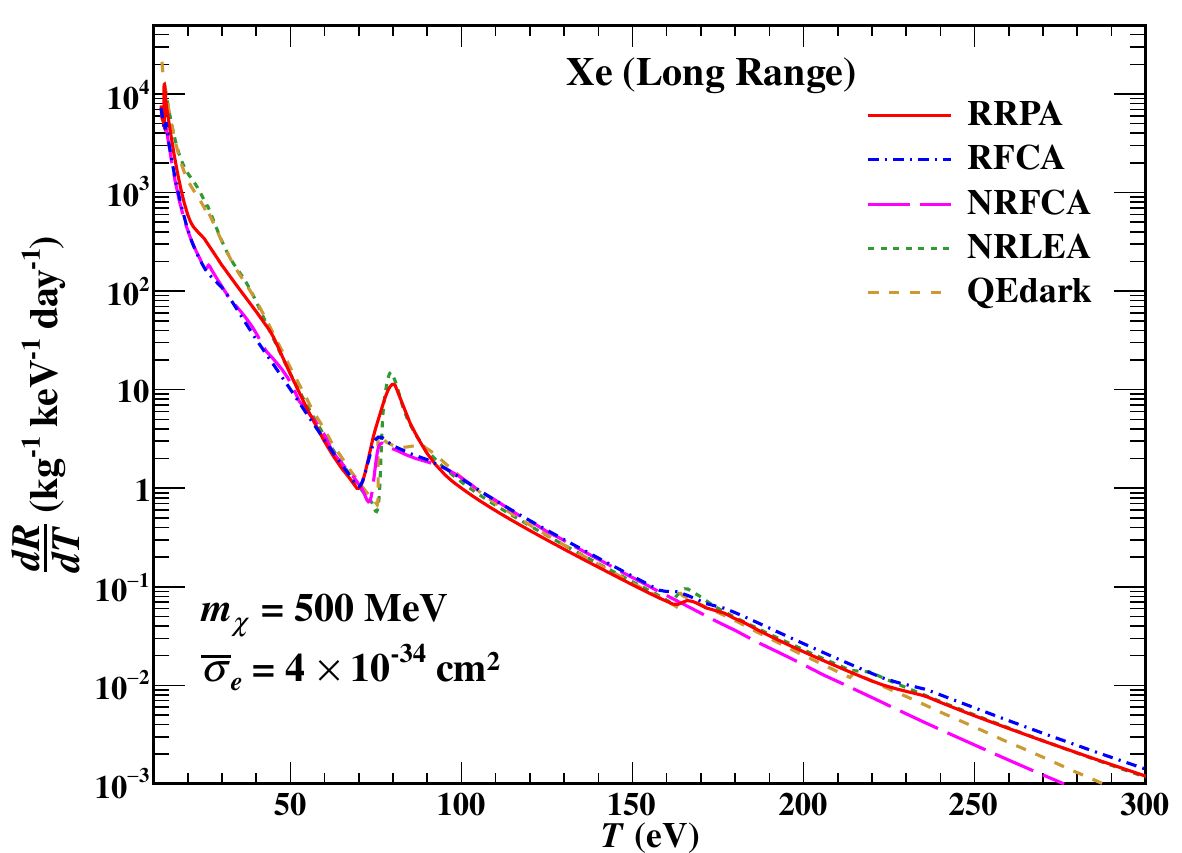}\tabularnewline
\end{tabular}
\par\end{centering}
\caption{Comparison of the differential rate as a function of energy transfer for a Xe atom, calculated using several models: this work (RRPA, red), relativistic FCA (blue), non-relativistic FCA (magenta), NRLEA~\citep{Hamaide:2021hlp}  (green), and the QEDark~\citep{Essig:2017kqs} (golden yellow). The calculations assume a dark matter mass
  $m_{\chi}$ =500 MeV and interaction strengths of (left) $c_{1} = 5.28 \times 10^{-4}$ and (right) $d_{1}=4.89 \times 10^{-11}$ (equivalent to ${\sigma}_{e}$ = $9 \times 10^{-42}$ $cm^{2}$ and $\bar{\sigma}_{e}$ = $4 \times 10^{-34}$ $cm^{2}$, respectively, in Ref.~\citep{Essig:2017kqs}).}
\label{fig:dR/dT:Comparisons}
\end{figure*}

In the Figure~\ref{fig:dR/dT:Comparisons}, we compare our RRPA with
a few previous works including our previous RFCA  and
nonrelativistic RFCA~\citep{Pandey:2018esq},
and two other nonrelativistic calculations Ref.~\citep{Hamaide:2021hlp}
(NRLEA, where LEA stands for local exchange approximation), and
Ref.~\citep{Essig:2017kqs} (QEDark),
reconstructed from the atomic ionization form factors provided in the
codes AtMolDM~\cite{AtMolDM} and QEdark~\cite{QEDarkCode} respectively.

Two important atomic ingredients: relativistic~\citep{Pandey:2018esq}
and exchange~\citep{Hamaide:2021hlp} effects have been discussed
previously. The former can be seen by the comparison of RFCA and NRFCA,
as the difference is caused solely by solving the Dirac and Schrödinger
equation, respectively, for the atomic wave functions. The latter
can be seen by the comparison of NRFCA and the NRLEA. While both approaches
are based on solving one-body Schrödinger equation for the final state,
the difference is on the formulations of the averaged central field
potential felt by the ionized electron. NRFCA only includes the direct
(Hartree) term, but NRLEA further adds an approximate, local exchange
potential. Even though the exchange potential is not formulated self-consistently,
the difference between NRLEA and NRFCA should still give some measure
of the exchange effect. Generally speaking, it enhances the event
rate at low energies ($T\lesssim 50~eV$), and suppresses at higher
energies ($T~> 100~eV$). 

With RRPA, we are able to incorporate the third important atomic ingredient:
the correlation effect, along with the relativistic and exchange effects,
in one self-consistent framework. From the comparison of the RRPA
and RFCA results, we see the combined exchange and correlation effect
enhances the event rate at low energies in 20$-$50 eV and suppresses at higher energies.
While not exactly the same as the previous NRLEA-NRFCA comparison (both have no correlation
effect), the general trend is quite consistent.

At last, we should point out an interesting consequence of combining
relativistic, exchange, and correlation effects in low-energy DM-electron
scattering. It was pointed in Refs.~\citep{Kopp:2009et,Chen:2015pha,Catena:2019gfa},
and explained in detail in Ref.~\citep{Liu:2021avx}, the leading-order
SD and SI DM-electron interactions can not be distinguished in DM
scattering off unpolarized targets, given that the atomic electrons
behaves as a non-relativistic, independent particle assembly. In this
case, the differential count rates of SI and SD scattering events
only differ by a constant factor of 3, so are not linearly independent.

\begin{figure*}
\begin{centering}
\begin{tabular}{cc}
\includegraphics[width=0.45\textwidth]{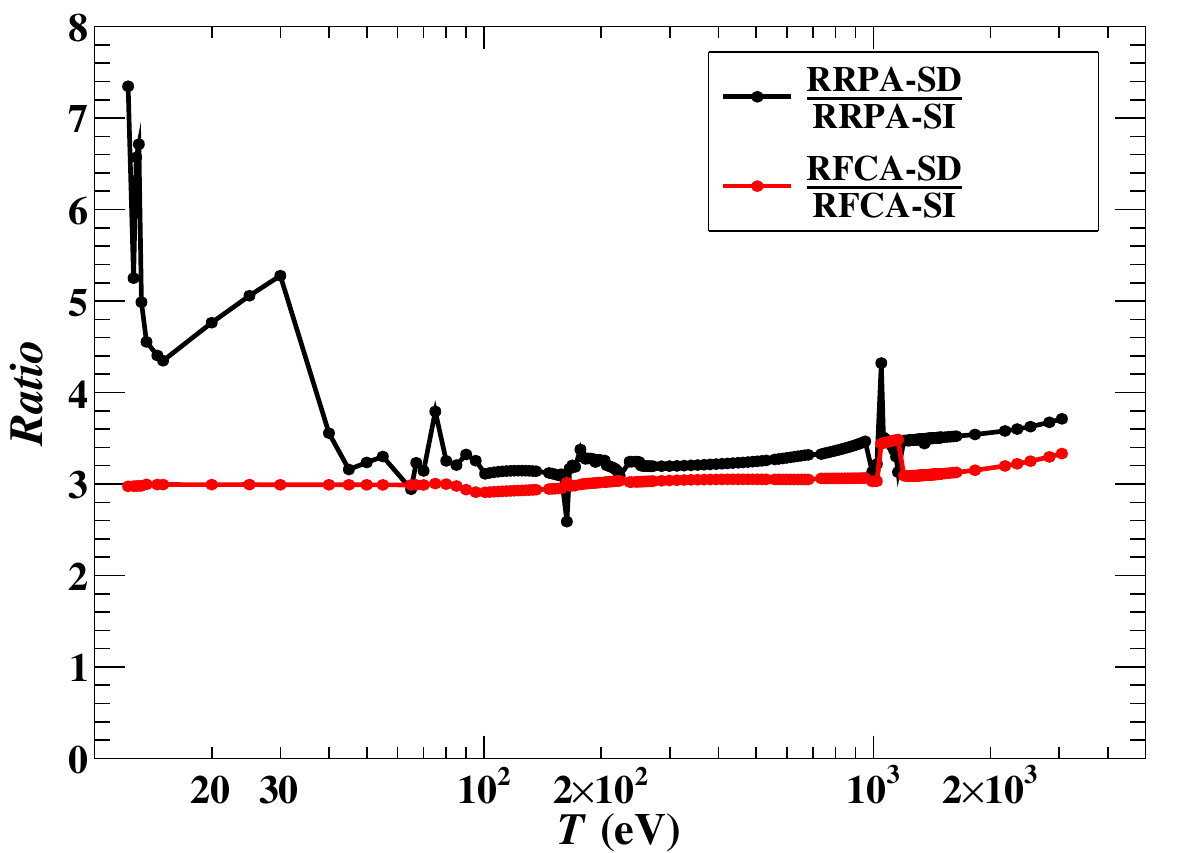}  & \includegraphics[width=0.45\textwidth]{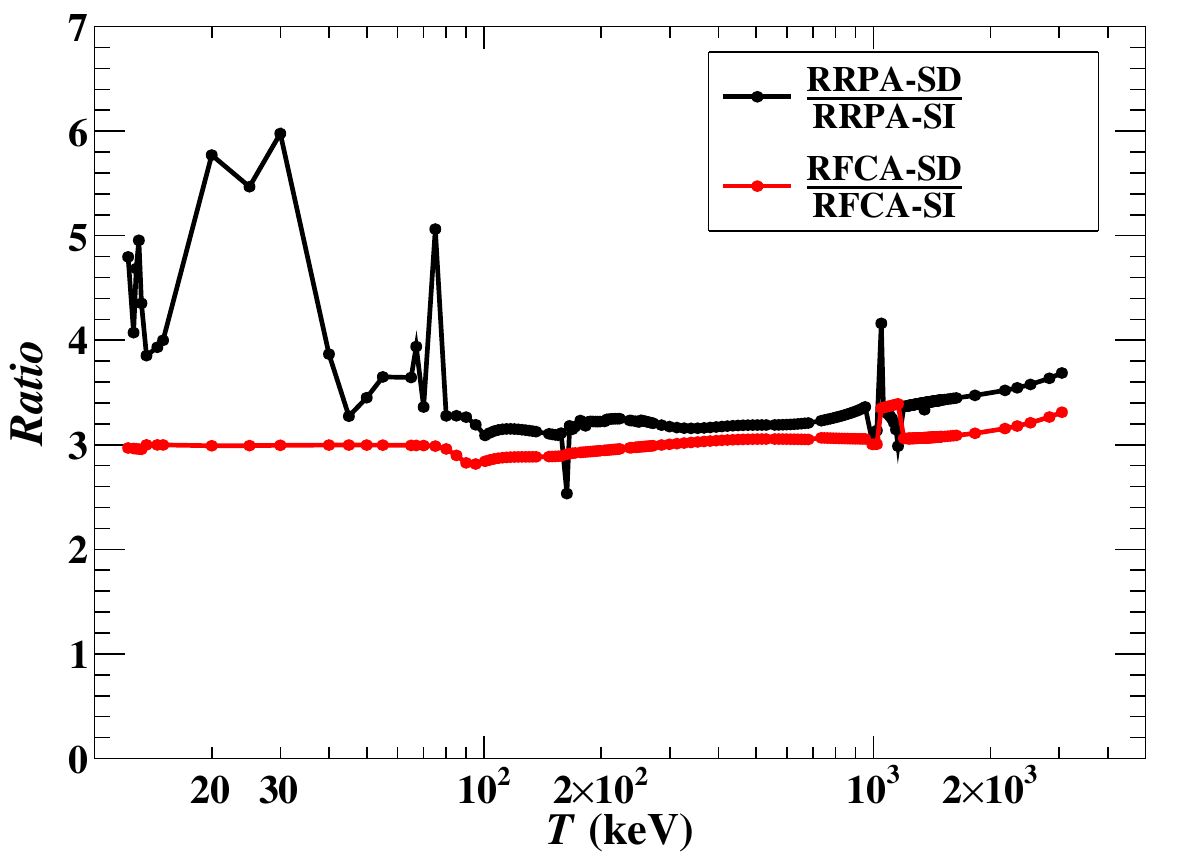}\tabularnewline
\end{tabular}
\par\end{centering}
\caption{Ratios of the differential rates due to the SD versus SI DM-electron
interactions of short- (left) and long- (right) range with $m_{\chi}=1\,\textrm{GeV}$.~}\label{fig:SDvsSI}
\end{figure*}

\begin{figure*}
\begin{centering}
\begin{tabular}{cc}
\includegraphics[width=0.45\textwidth]{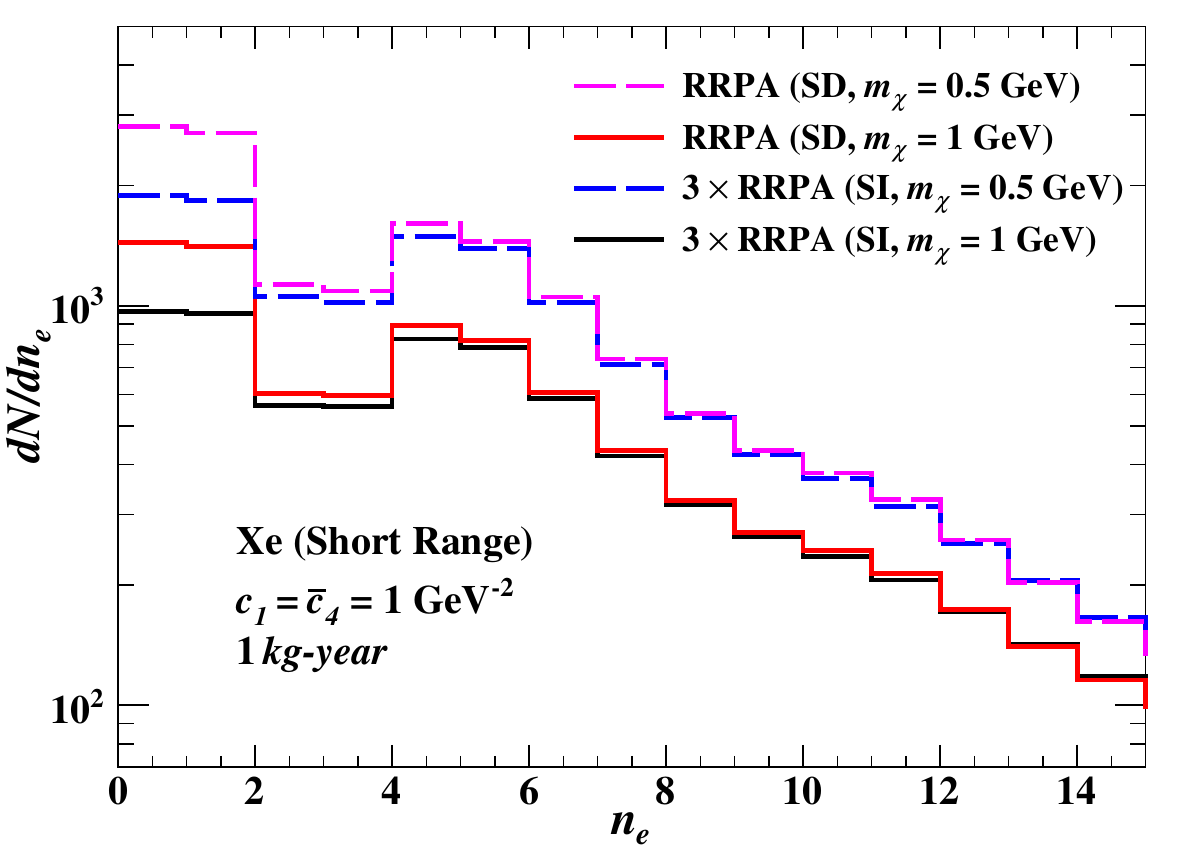}  & \includegraphics[width=0.45\textwidth]{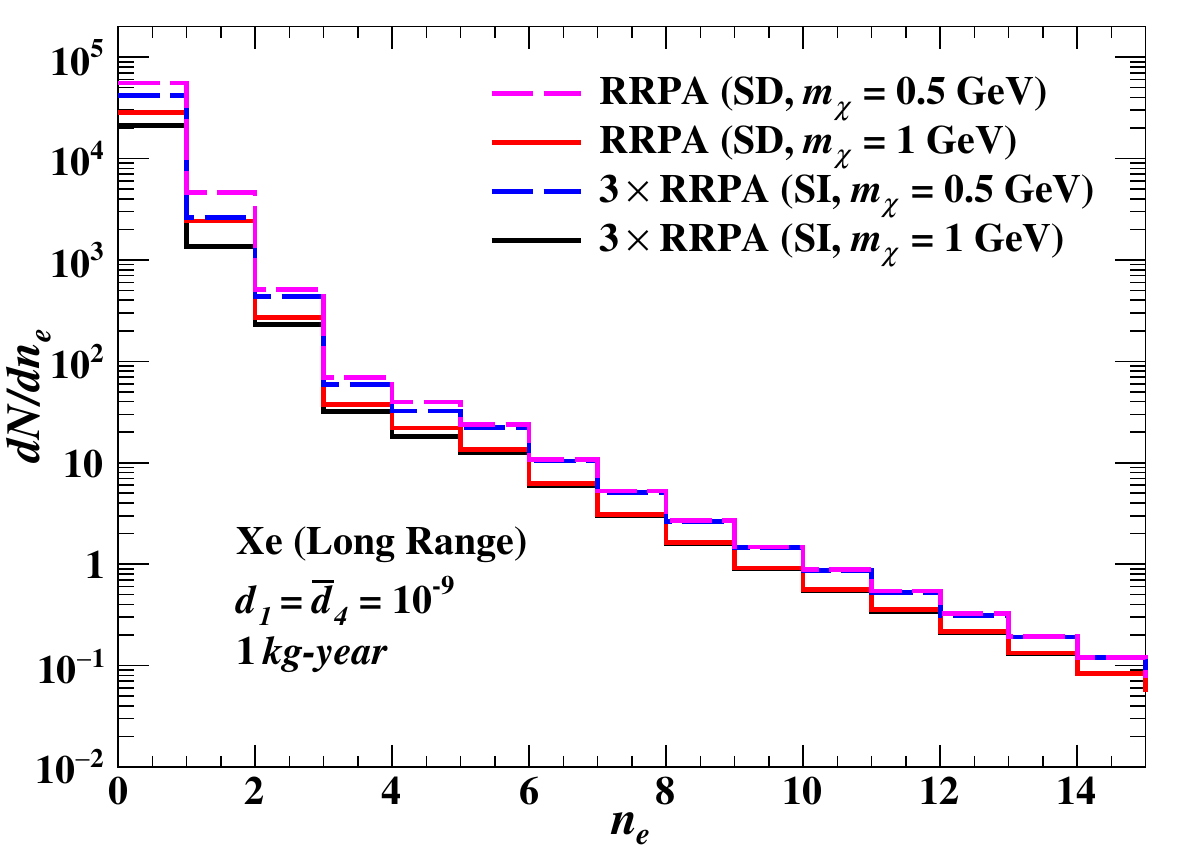}\tabularnewline
\end{tabular}
\par\end{centering}
\caption{Comparisons of the expected event numbers due to the SD and SI DM-electron
short-(left) and long-ranged (right) interactions as a function of
the ionized electron number. The SI event numbers are multiplied by
the scaling factor of 3, such that differences of the SD versus SI
spectral shapes can be clearly seen.~}\label{fig:dNe/dEe_SDvsSI}
\end{figure*}

As shown in Fig.~\ref{fig:SDvsSI}, the SD-SI ratio of $dR/dT$ by
RFCA has deviation from the factor of 3 but only in $T\gtrsim1\,\textrm{keV}$.~\footnote{We note a coding mistake was identified in the work of Ref.~\citep{Liu:2021avx}
that underestimates the relativistic effect in the SD part. An erratum
is in preparation. } It can be attributed to the fact that the spin-orbit interaction,
as a part of the relativistic effect, is more pronounced for inner-shell
electrons whose ionizations dominate at high energy transfer. With
RRPA further including the exchange and correlation effects, we found
the SD-SI ratio deviates from 3 substantially at low energies of $T\lesssim100\,\textrm{eV}$.
This implies that the SD DM-electron interaction can possibly be differentiated
from the SI one, and treated as an independent component of DM-electron
interactions in conventional DM direct searches with unpolarized detector
media. 

While we will defer further theory explanations for future work, we
can comment that such enhanced spin-dependent effect due to relativistic
many-body physics at low $T$ is not unexpected. For ground-state
xenon, the spin-orbit splitting of the two most outer shells $E_{5p_{3/2}}-E_{5p_{1/2}}=1.3\,\textrm{eV}$,
which is not small: $\sim10\%$ of the first ionization energy $-E_{5p_{3/2}}=12.1\,\textrm{eV}$.
For ground-state atomic germanium, while $E_{4p_{1/2}}-E_{4p_{3/2}}=0.2\,\textrm{eV}$
seems smaller at $\sim3\%$ of the first ionization energy $-E_{4p_{1/2}}=7.8\,\textrm{eV}$,
one should take note that the ordering of $4p_{1/2}$ and $4p_{3/2}$
orbitals is reversed, in contrary to the usual Landé interval rule.

In Fig.~\ref{fig:dNe/dEe_SDvsSI}, we convert the differential rate ($dR/dT$)
into the expected event numbers as a function of ionized
electron number ($dN/dn_{e}$) through the following procedure. First, the recoil energy of the primary electron,
denoted as $E_{r}$, is calculated by subtracting the average energy
required to create one free quantum in liquid xenon ($W$) from the total energy transferred ($T$):
\begin{equation}
  E_{r}~=~T - W.
\end{equation}
Recent measurements indicate that the value for $W$ is 11.5 eV~\cite{EXO-200:2019bbx, Baudis:2021dsq}.
The maximum number of electrons ($N_{max}$) which include the primary and additional quanta from de-exciation, is  obtained from the charge yield ($Q_{y}$) measured in~\cite{XENON:2021qze}  as
\begin{equation}
  N_{max}~=~E_{r}.Q_{y}.
\end{equation}
The expected event rate is evaluated from the differential recoil spectrum $dR/dE_{r}$ using the following euqation,  
\begin{equation}
\frac{dN}{dn_{e}} = \int_{E_{r}^{min}}^{E_{r}^{max}} \frac{dR}{dE_{r}} Binomial(n_{e};N_{max}; f_{e}) dE_{r}.
\end{equation}
where $n_{e}$ follows a binomial distribution with $N_{max}$ trials and success proability of $f_{e}$ = 0.83~\cite{Essig:2017kqs}. Note that our approach of converting $dR/dT$ to $dN/dn_{e}$ differs slightly from Ref~\cite{Essig:2017kqs}. 

The expected event numbers as a function of ionized
electron number $n_{e}$, assuming a xenon detector with 1 kg-year
exposure and a $1$-GeV DM particle is illustrated in Figure~\ref{fig:dNe/dEe_SDvsSI}.
One can easily observes their differences in spectral
shapes in comparison to the corresponding SI ones multiplied by the
factor of 3. This definitely will enrich the direct search data analyses
focusing at low energies, for example, the recent XENON1T SE data
set~\citep{XENON:2021qze}.

\subsection*{New exclusion limits}

Dual-phase liquid xenon detectors have emerged as a vital technology
in the search for dark matter, particularly for light dark matter
(LDM), due to their exceptional sensitivity to ionization events down
to the level of a single electron. This remarkable ability to detect
single-electron ionization has greatly advanced the study of dark
matter interactions, especially those involving electron scattering.
Several key experiments using these detectors, such as XENON10, XENON100,
XENON1T, PANDAX-II, and XENONnT, have made substantial contributions
to the growing body of data in dark matter research.

Each of these experiments focuses on detecting dark matter by observing
its possible interactions with atomic nuclei or electrons within the
xenon target material. To explore and set exclusion limits on SI or
SD DM-electron interactions and dark matter mass, data from experiments including
XENON10~\cite{XENON10:2011prx}, XENON100~\cite{XENON:2016jmt},
XENON1T~\cite{XENON:2019gfn}, and PANDAX-II~\cite{PandaX-II:2021nsg}
are analyzed in comparison with theoretical differential rate
calculated by our RRPA response functions. 

\begin{figure}
\begin{centering}
\begin{tabular}{cc}
\includegraphics[width=0.45\textwidth]{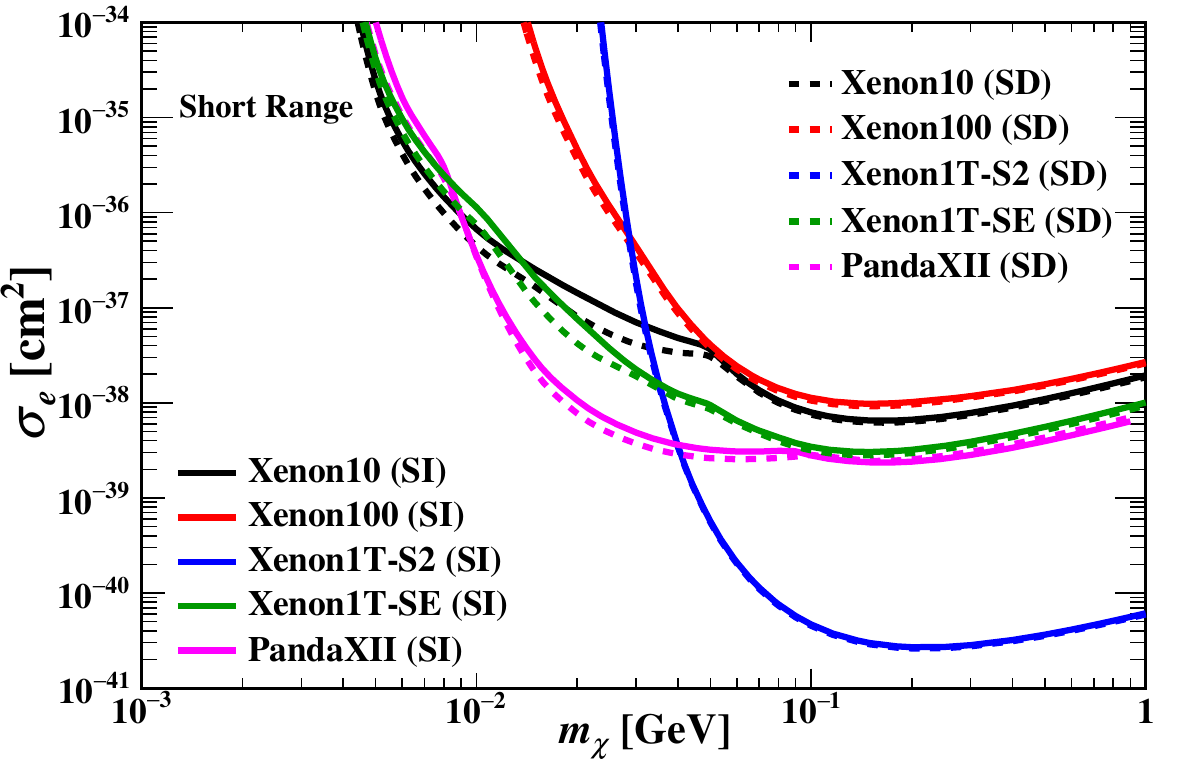}  & \includegraphics[width=0.45\textwidth]{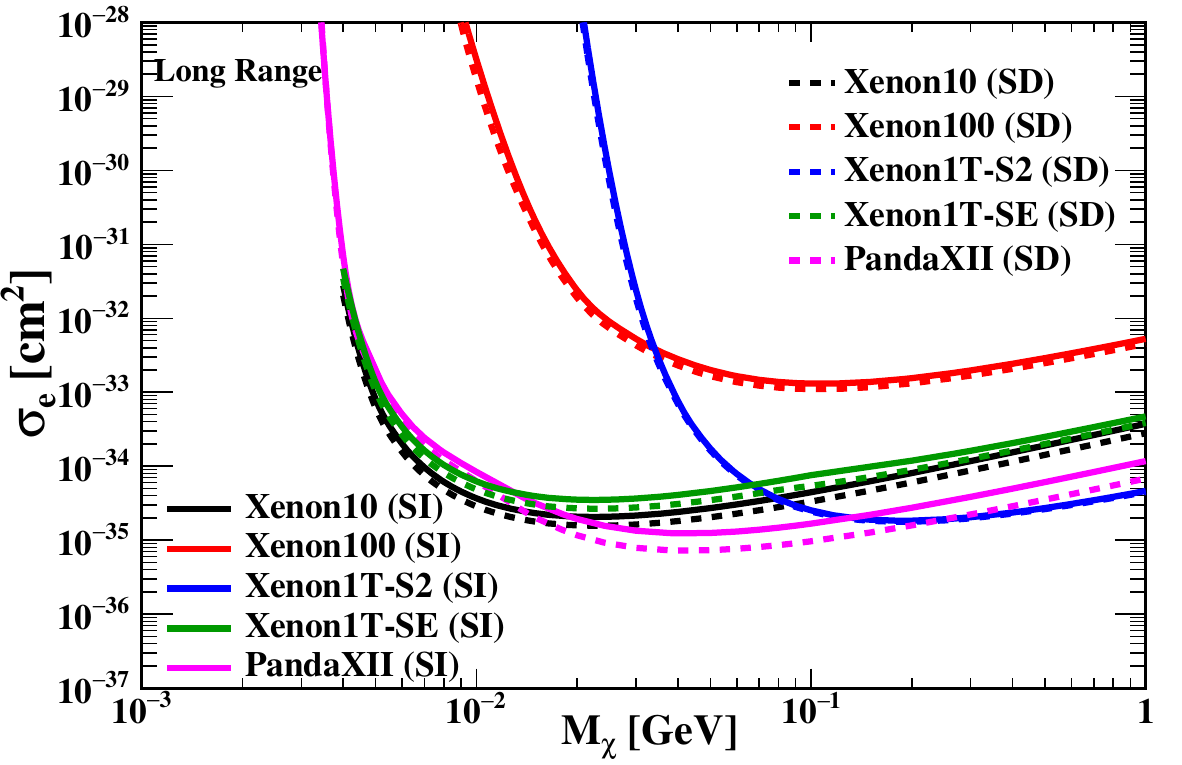}\tabularnewline
\end{tabular}
\par\end{centering}
\caption{The exclusion limits at 90\% confidence level (C.L.) on spin-independent
and spin-dependent short-range (left panel) and long-range (right
panel) DM-electron interactions, as functions of dark matter mass,
are derived from XENON experiment data, including XENON10 (black)~\cite{Essig:2012yx},
XENON100 (red)~\cite{Essig:2017kqs}, XENON1T-S2 (blue)~\cite{XENON:2019gfn},
XENON1T-S2 (green)~\cite{XENON:2021qze}, and PandaX-II~\cite{PandaX-II:2021nsg}.~}\label{fig:exclusion}
\end{figure}

Under a conservative assumption that all observed events could be
attributed to potential DM-electron scattering, upper limits at a
90\% confidence level (C.L.) have been derived for both short- and
long-range interactions, following the same methodology as outlined
in previous studies, such as Refs.~\cite{Pandey:2018esq,Liu:2021avx}.
These limits are shown in Fig.~\ref{fig:exclusion}. The DM-free-
electron scattering cross section is $\sigma_{e}=c_{1}^{2}\mu_{\chi e}^{2}/\pi$
or $\sigma_{e}=d_{1}^{2}\mu_{\chi e}^{2}/(\pi m_{e}^{4}\alpha^{4})$
for the SI SR or LR interaction, respectively. For the SD interaction,
the coupling constant squared is changed to $3\bar{c}_{4}^{2}$ or
$3\bar{d}_{4}^{2}$, accordingly. In the case of XENON1T, the data
from both the XENON1T S2-only and the XENON1T single-electron (SE)
data set have been analyzed. The XENON1T-SE~\cite{XENON:2019gfn}
data set has a lower threshold, which allows it to probe lower dark
matter masses, but near the threshold the background levels becomes
higher. As a result, while the XENON1T-SE~\cite{XENON:2021qze}
data can explore lower mass ranges, the constraints degrade at higher
masses due to the increased background noise.

For both the SR and LR SI interactions, the limits set by the XENON1T-S2
data, which has a relatively high-threshold at $186\,\textrm{eV}$,
become slightly less stringent from RFCA to RRPA. This can be seen
in Fig.~\ref{fig:diff_RRPA/FCA} that RFCA predicts a slightly bigger
count rate for most region in $T\gtrsim186\,\textrm{eV}$. On the
other hand, with the low-threshold XENON10 data, at $1\,n_{e}$ (XENON1T-SE
and PandaXII alike), the situation is more subtle. As Fig.~\ref{fig:diff_RRPA/FCA}
shows, the energy region that RRPA differs from RFCA is $T\lesssim300\,\textrm{eV}$,
and predicts a bigger count rate in $T\lesssim50\,\textrm{eV}$. As
a result, the exclusion limits on the LR SI interaction or the SR
interaction with a low-mass DM particle are improved from RFCA to
RRPA, since in both cases the scattering events are dominated by low
energy bins.

While similar reasoning applies the the case of the SD interactions,
one should first note that the scaling factor of $3$ is also built-in
in the DM scattering with a free electron, i.e. $\sigma_{e}^{(\textrm{SD})}=3\sigma_{e}^{(\textrm{SI})}$,
given the same coupling constants $c_{1}=\bar{c}_{4}$ for the SR,
or $d_{1}=\bar{d}_{4}$ for the LR interaction. If the unpolarized
atomic response functions follow the same scaling, the exclusion limit
on $\sigma_{e}^{(\textrm{SD})}$ would be exactly the same as the
one on $\sigma_{e}^{(\textrm{SI})}$. This is what can be seen in
Fig.~\ref{fig:exclusion}: The XENON1T-S2 exclusion curves for the
SD (in dashed blue) interactions can barely be distinguished from
the ones for the SI counterparts (in solid blue). As one sees in Fig.~\ref{fig:SDvsSI},
the deviation from the scaling factor is modest in $T\gtrsim186\,\textrm{eV}$,
and is only sizable in the energy range where the count rate is small.
However, for data sets with low energy thresholds, the substantial
scaling deviation ($T\lesssim300\,\textrm{eV}$ in Fig.~\ref{fig:SDvsSI})
result in not only a energy spectral shape different from the SI case,
but also excess counts (see Fig.~\ref{fig:dNe/dEe_SDvsSI}) that
give rise to the more stringent exclusion limits than the SI interactions.

\section{Summary~}\label{sec:summary}

In this work, we applied a state-of-the-arts atomic technique, the
(multiconfiguration) relativistic random phase approximation, to compute
the response functions of sub-GeV light dark matter particles scattered
off the xenon and germanium atoms through either spin-independent
or spin-dependent dark matter-electron interactions at leading order
in effective field theory expansion. Our method, unlike most existing
atomic approaches applied to sub-GeV dark matter searches, is successfully
benchmarked by photoabsorption data in the energy range of tens of
eV to 30 keV, which covers the kinetic energy range of dark matter
particles with masses in between MeV and GeV. In addition to previously-identified
contributions from relativistic and exchange effects, the correlation
effect not included in independent particle frameworks is incorporated
by RPA. Surprisingly, these effects combined to cause different energy
dependence of the spin-dependent versus the spin-independent responses,
such that two interactions can be distinguished at low energies, unlike
previous results and argument based on the independent particle picture.
Lastly, we use these high-quality response functions to place reliable
exclusion limits with data from current experiments including XENON10,
XENON100, XENON1T, PANDAX-II, and XENONnT.

The response functions computed in this work are available for download.
The codes that automatize the process of getting detector count rate
predictions are also provided. Our future plan includes extending
the parameter space of response functions, decomposing the response
functions by individual atomic shells, and including other atomic
species that serve as detector materials. 
\begin{acknowledgments}
This work was supported in part under Contract Nos. 111-2112-M259-007
(C.-P. L.), 111-2811-M-259-014 (M. K. P.), from the Ministry of Science
and Technology; 112-2112-M259-014 (C.-P. L.), 112-2811-M-259-006 (C.-P.
W.), 112-2112-M-002-027 and 113-2112-M-002-012 (J.-W. C.), 113-2112-M-001-053-MY3
(H. T. W.) from the National Science and Technology Council; and 2021-2024/TG2.1
from the National Center for Theoretical Sciences of Taiwan; the Canada
First Research Excellence Fund through the Arthur B. McDonald Canadian
Astroparticle Physics Research Institute (C.-P. W.); and Contract
F.30-584/2021 (BSR), UGC-BSR Research Start Up Grant, India;
the PURSE grant  and FIST program of Department of Science \& Technology,
India (L. S.).
We thank the Academia Sinica Grid-computing Centre for providing the
computing facility that makes this work possible, and its staff for
excellent technical supports. JWC thanks the InQubator for Quantum
Simulation in Seattle and the Yukawa Institute for Theoretical Physics
at Kyoto Univ. for hospitality. 
\end{acknowledgments}

\appendix*

\section{Brief Overview of (MC)RRPA Theory}

While there are many ways to formulate the (MC)RRPA theory, in this
appendix we follow the route of time-dependent Dirac-Hartree-Fock
theory (the relativistic version of time-dependent Hartree-Fock),
as it becomes particularly handy when it comes the computation of
the many-body wave functions and transition amplitudes. 

\subsection*{General Formalism}

Consider a time-dependent Hamiltonian of an atomic system with \textit{N}
electrons given by: 
\begin{equation}
H(t)=H_{0}+V(t)\,.
\end{equation}
The time-independent Hamiltonian 
\begin{align}
H_{0} & =\sum_{i=1}^{N}h_{i}+{\displaystyle \sum_{i<j}^{N}\dfrac{e^{2}}{r_{ij}}\,,}\\
h_{i} & =\vec{\alpha}_{i}\cdot\vec{p}_{i}+\beta m_{e}-\frac{Z}{r_{i}}\,,
\end{align}
governs the usual atomic structure with the Dirac Hamiltonian, nuclear
Coulomb attraction, and electron-electron repulsion. The time-dependent
interaction that causes atomic transitions is assumed to take a one-body
harmonic form 
\begin{equation}
V(t)=\sum_{i=1}^{N}v_{i}e^{-i\omega t}+\sum_{i=1}^{N}v_{i}^{\dagger}e^{+i\omega t}\,\equiv V_{+}e^{-i\omega t}+V_{-}e^{i\omega t}\,,\label{eq:V(t)}
\end{equation}
It will be treated perturbatively in the end, and become clear that
the $v_{i}(v_{i}^{\dagger})$ term results in a single electron (de-)excitation
of energy $\omega$. In this work, $V(t)$ is the DM-electron interaction
due to Eq.~(\ref{eq:L_EFT}). 

To solve the many-body wave function $\ket{\Psi(t)}$, the equation
of motion 
\begin{equation}
i\frac{\partial}{\partial t}\ket{\Psi(t)}=H(t)\ket{\Psi(t)}\,,
\end{equation}
can be recast into a Frenkel variational form:
\begin{equation}
\left<\delta\Psi(t)\left\vert \left[i\frac{\partial}{\partial t}-H(t)\right]\right\vert \Psi(t)\right>=0\,.\label{eq:TDHF-1}
\end{equation}
By further factoring out the time evolution due to the unperturbed
eigenenergy, $E$, i.e., 
\begin{equation}
\ket{\Psi(t)}=e^{-iEt}\ket{\Psi^{'}(t)}\,,
\end{equation}
the variational solution of $\ket{\Psi^{'}(t)}$ is 
\begin{equation}
\left<\delta\Psi^{'}(t)\left\vert \left[E+i\frac{\partial}{\partial t}-H(t)\right]\right\vert \Psi^{'}(t)\right>=0\,.\label{eq:TDHF-2}
\end{equation}

In the spirit of Hartree-Fock (HF) theory, we seek an approximate
solution to Eq.~(\ref{eq:TDHF-2}) based on Slater determinants with
relativistic Hamiltonian, referred as Dirac-Fock (DF) theory
henceforth. First, $\ket{\Psi^{'}(t)}$ is formed by a linear combination
of configuration state functions (CSFs) $\ket{\Phi_{A}(t)}$ 

\begin{equation}
\ket{\Psi^{'}(t)}={\displaystyle \sum_{A}C_{A}(t)\ket{\Phi_{A}(t)}\,,}\label{eq:TDHF_wf}
\end{equation}
where $A$ is the configuration index, $C_{A}(t)$ the time-dependent
weight. Note that the multi-configuration (MC) feature is simply reflected
by the number of possible $A$'s is more than one. A configuration
$A$ for a $N$-electron state is specified by the $N$ electron orbitals
being occupied and labeled by $\{\alpha_{1},\,\alpha_{2},\,\ldots\alpha_{N}\}$
with each $\alpha$ collectively denotes four quantum numbers: principle
$n_{\alpha}$ (or reduced wave number $k_{\alpha}$ for a continuum
state), orbital and total angular momentum $l_{\alpha}$ and $j_{\alpha}$
(in relativistic cases, they are usually combined as $\kappa_{\alpha}$,
which is $\mp(j_{\alpha}+1/2)$ for $l_{\alpha}=j_{\alpha}\pm1/2$)),
and the $z$-axis projection of total angular momentum $m_{\alpha}$.
The CSFs are taken to be the Slater determinant of all occupied time-dependent single electron orbitals, denoted by 

\begin{equation}
\ket{u_{\alpha}(t)}=\ket{n_{\alpha}l_{\alpha}j_{\alpha}m_{\alpha};t}\equiv\ket{am_{\alpha};t}\,.
\end{equation}

The equation of motions for $C_{A}(t)$'s and $u_{\alpha}(t)$'s are
derived by varying $C_{A}^{*}(t)$'s and $u_{\alpha}^{\dagger}(t)$
in Eq.~(\ref{eq:TDHF-2}) with the constraints of probability conservation
\begin{equation}
\left<\Psi^{'}(t)\vert\Psi^{'}(t)\right>=1\,.\label{eq:wf_norm}
\end{equation}
This can be furnished by the following conditions 
\begin{align}
\sum_{A}C_{A}^{\star}(t)C_{A}(t) & =1\,,\label{eq:constraints_CA}\\
\left<u_{\alpha}(t)\vert u_{\beta}(t)\right> & =\delta_{\alpha\beta}\,,\label{eq:constraints_ualpha}
\end{align}
where orbital wave functions are required to be orthonormal at all
times, and configuration weights satisfy the normalization condition.
This leads to 
\begin{align}
(E+i\frac{\partial}{\partial t})C_{A}(t)+\sum_{B}\left\langle \Phi_{A}(t)\left|i\frac{\partial}{\partial t}-H(t)\right|\Phi_{B}(t)\right\rangle  & =0\,,\\
\sum_{AB}C_{A}^{*}(t)C_{B}(t)\frac{\partial}{\partial u_{\alpha}^{\dagger}(t)}\left\langle \Phi_{A}(t)\left|i\frac{\partial}{\partial t}-H(t)\right|\Phi_{B}(t)\right\rangle  & =\sum_{\beta}\gamma_{\alpha\beta}(t)u_{\beta}(t)\,,
\end{align}
where $\partial/\partial u_{\alpha}^{\dagger}(t)$ should be understood
as a functional derivative acting on the many-body matrix element
that follows, and $\gamma_{\alpha\beta}(t)$ the Lagrange multiplier
that enforces the orthonormality of single-particle orbitals. 

As the time-dependent interaction $V(t)$ is harmonic in time, the
solutions of $C_{A}(t)$, $u_{\alpha}(t)$, and $\gamma_{\alpha\beta}(t)$
can assume the following harmonic expansions with time dependence
explicitly factored out by $e^{\pm in\omega t}$: 
\begin{equation}
C_{A}(t)=C_{A}+\left[C_{A}\right]_{+}e^{-i\omega t}+\left[C_{A}\right]_{-}e^{+i\omega t}+\ldots,\label{eq:RPA_C}
\end{equation}
\begin{align}
u_{\alpha}(t) & =u_{\alpha}+w_{\alpha+}e^{-i\omega t}+w_{\alpha-}e^{+i\omega t}+\ldots,\label{eq:RPA_u}\\
\gamma_{\alpha\beta}(t) & =\gamma_{\alpha\beta}+\left[\gamma_{\alpha\beta}\right]_{+}e^{-i\omega t}+\left[\gamma_{\alpha\beta}\right]_{-}e^{+i\omega t}+\ldots,\label{eq:RPA_gamma}
\end{align}
where ``$\ldots$'' denotes higher harmonic terms. The so-called
random phase approximation (RPA) refers to truncation of the series
to the first harmonic terms with all ``$\ldots$'' terms omitted.
Viewing $V(t)$ as a perturbation, RPA corresponds to a time-dependent
perturbation theory to first order, so sometimes is also called the
linear response theory. The time-independent configuration weights,
$C_{A}$ and $\left[C_{A}\right]_{\pm}$, and orbital wave functions
$u_{\alpha}$ and $w_{\alpha\pm}$ are now subject to the following
constraints 
\begin{align}
 & \sum_{A}C_{A}^{*}C_{A}=1\,\\
 & \sum_{A}\left[C_{A}\right]_{\mp}^{*}C_{A}+C_{A}^{*}\left[C_{A}\right]_{\pm}=0\,,\\
 & \braket{u_{\alpha}|u_{\beta}}=\delta_{\alpha\beta}\,,\\
 & \braket{w_{\alpha\mp}|u_{\beta}}+\braket{u_{\alpha}|w_{\alpha\pm}}=0\,.
\end{align}

Because the perturbation is a one-body operator, one thing it can
do is promoting an electron in a filled orbital, $u_{\alpha}$, to
an unoccupied state (in our case in continuum). As the energy increases
(manifested by $e^{-i\omega t}$), this state is labeled by $w_{\alpha+}$
and can be expanded by unoccupied orbitals. A calculation that only
takes into these states is called Tamm-Dancoff approximation (TDA).
The improvement of RPA over TDA is introducing the correlation effect
to the ground state wave function, such that it contains other CSFs
than the ones in the MCDF reference state. As a result, the other
thing that the one-body operator can do is demoting an electron from
a filled orbital of higher energies back to a vacancy in $u_{\alpha}$.
As the energy decreases (manifested by $e^{+i\omega t}$), this state
is labeled by $w_{\alpha-}$. Therefore, RPA incorporates the correlation
effects in the ground and excited states with equal footing. 

Using the harmonic expansion above, at zeroth order, one obtains the
MCDF equations for unperturbed weights $C_{A}$ and orbitals $u_{\alpha}$:
\begin{align}
EC_{A}-\sum_{B}\left\langle \Phi_{A}\left|H_{0}\right|\Phi_{B}\right\rangle C_{B} & =0\,,\\
\sum_{AB}C_{A}^{*}C_{B}\frac{\partial}{\partial u_{\alpha}^{\dagger}}\left\langle \Phi_{A}\left|H_{0}\right|\Phi_{B}\right\rangle  & =\sum_{\beta}\gamma_{\alpha\beta}u_{\beta}\,.
\end{align}
At first order, one obtains the MCRRPA equations for $\left[C_{A}\right]_{\pm}$
and $w_{\alpha\pm}$:

\begin{align}
 & (E\pm\omega)\left[C_{A}\right]_{\pm}-\sum_{B}\left(\left\langle \Phi_{A}\left|H_{0}\right|\Phi_{B}\right\rangle \left[C_{B}\right]_{\pm}+\left\langle \Phi_{A}\left|H_{0}\right|\Phi_{B}\right\rangle _{\pm}C_{B}\right)\nonumber \\
 & =\sum_{B}\left\langle \Phi_{A}\left|V_{\pm}\right|\Phi_{B}\right\rangle C_{B}\,,
\end{align}

\begin{align}
 & \sum_{AB}C_{A}^{*}C_{B}\frac{\partial}{\partial u_{\alpha}^{\dagger}}\left\langle \Phi_{A}\left|i\frac{\partial}{\partial t}-H_{0}\right|\Phi_{B}\right\rangle _{\pm}-\sum_{AB}\left(\left[C_{A}\right]_{\mp}^{*}C_{B}+C_{A}^{*}\left[C_{B}\right]_{\pm}\right)\frac{\partial}{\partial u_{\alpha}^{\dagger}}\left\langle \Phi_{A}\left|H_{0}\right|\Phi_{B}\right\rangle \nonumber \\
 & -\sum_{\beta}\gamma_{\alpha\beta}w_{\beta\pm}+\left[\gamma_{\alpha\beta}\right]_{\pm}u_{\beta}=\sum_{AB}C_{A}^{*}C_{B}\frac{\partial}{\partial u_{\alpha}^{\dagger}}\left\langle \Phi_{A}\left|V_{\pm}\right|\Phi_{B}\right\rangle \,.
\end{align}
Note that the energy functional $\left\langle \Phi_{A}\left|i\frac{\partial}{\partial t}-H_{0}\right|\Phi_{B}\right\rangle _{\pm}$
contains the perturbed wave function $w_{\alpha\pm}$ in the ket state
$\ket{\Psi_{B}}$ or $w_{\alpha\mp}^{\dagger}$ in the bra state $\bra{\Psi_{A}}$.

\subsection*{Applications to Ionization Processes of Xenon and Germanium}

Up to this stage, the MCDF and MCRRPA equations are completely general,
and can involve an arbitrary number of CSFs. To bring the numerical
computations to a manageable scale, one has to limit the numbers of
CSFs and the single electron orbitals on which they are built. As
we deal in this work the problem of atomic ionization by one-body
perturbations caused by DM-electron interactions, the most relevant
CSFs are the ones that describe the atomic ground state, and the one-particle-one-hole
excitations from the ground state that describe the ion-plus-one-free-electron
final states. 

For xenon, a closed-shell atom, we follow the conventional HF approach
and treat its ground state wave function at zeroth order, the reference
state, as a single Slater determinant with the configuration $[\textrm{Kr}]4d^{10}5s^{2}5p^{6}$,
and this gives the set of the unperturbed orbitals $u_{\alpha}$'s.
As all orbitals are completely filled, its total angular momentum
and parity is $0^{+}$. For germanium, a divalent atom, its ground
state configuration at zeroth order is $[\textrm{Ar}]3d^{10}4s^{2}4p^{2}$
with total angular momentum and parity being $0^{+}$. In this case,
the reference state contains two CSFs: one with $4p_{3/2}^{2}$, the
other with $4p_{1/2}^{2}$. Both can form a $0^{+}$ state, but are
made of different $4p$ orbitals. For this distinction, the computations
of xenon and germanium are labeled as RRPA and MCRRPA, respectively.

With these considerations, the above MCDF and MCRRPA equations can
be made more explicitly. Here we use the case of xenon as an example.
Because the reference state has only one configuration, there is no
equation for $C_{A}$'s to be solved (only $C_{1}=1$). The orbital
equation at zeroth oder becomes:
\begin{equation}
\left[h+v_{12}^{(\textrm{DF})}\right]u_{\alpha}(\vec{r})=\epsilon_{\alpha}u_{\alpha}(\vec{r})+\sum_{\beta\neq\alpha}\gamma_{\alpha\beta}u_{\beta}(\vec{r})\,,\label{eq:DF_orb_eqn}
\end{equation}
with 
\begin{equation}
v_{12}^{(\textrm{DF})}u_{\alpha}(\vec{r})=\sum_{\beta}\int d^{3}r^{'}\,\frac{e^{2}}{|\vec{r}-\vec{r}'|}\left[u_{\beta}^{\dagger}(\vec{r}^{'},t)u_{\beta}(\vec{r}^{'},t)u_{\alpha}(\vec{r},t)-u_{\beta}^{\dagger}(\vec{r}^{'},t)u_{\beta}(\vec{r},t)u_{\alpha}(\vec{r}^{'},t)\right]\,.\label{eq:DF_pot}
\end{equation}
The Dirac-Fock (DF) potential is simply a relativistic version of
the conventional HF potential which consists of the direct and exchange
terms, and the diagonal Lagrange multiplier $\gamma_{\alpha\alpha}$
is identified as the orbital energy $\epsilon_{\alpha}$. The orbital
equation at first order becomes 
\begin{equation}
\left[h+v_{12}^{(\textrm{RPA})}\right]w_{\alpha\pm}(\vec{r})=(\epsilon_{\alpha}\pm\omega)w_{\alpha\pm}(\vec{r})+v_{\pm}u_{\alpha}(\vec{r})+\sum_{\beta}\left[\gamma_{\alpha\beta}\right]_{\pm}u_{\beta}(\vec{r})+\sum_{\beta\neq\alpha}\gamma_{\alpha\beta}w_{\beta\pm}(\vec{r})\,,\label{eq:RRPA_orb_eqn}
\end{equation}
with
\begin{align}
v_{12}^{(\textrm{RPA})}w_{\alpha\pm}(\vec{r}) & =v_{12}^{(\textrm{DF})}w_{\alpha\pm}(\vec{r})+\sum_{\beta}\int d^{3}r^{'}\,\frac{e^{2}}{|\vec{r}-\vec{r}'|}\Big[u_{\beta}^{\dagger}(\vec{r}^{'})w_{\beta\pm}(\vec{r}^{'})u_{\alpha}(\vec{r})\nonumber \\
 & -u_{\beta}^{\dagger}(\vec{r}^{'})w_{\beta\pm}(\vec{r})u_{\alpha}(\vec{r}^{'})w_{\beta\mp}^{\dagger}(\vec{r}^{'})u_{\beta}(\vec{r}^{'})u_{\alpha}(\vec{r})-w_{\beta\mp}^{\dagger}(\vec{r}^{'})u_{\beta}(\vec{r})u_{\alpha}(\vec{r}^{'})\Big]\,.\label{eq:RRPA_pot}
\end{align}

Moving on to open-shell atoms, the algebra becomes more cumbersome
as the numbers of valence electrons increase. For divalent systems
of zero total angular momenta, the details can be found in Ref.~\cite{Huang:1981wj}. 

After the zeroth-order ground state configurations are fixed, the
excited state configurations are determined by the transition operators
prescribed by the perturbation. As rotation and parity are good symmetries
of $H_{0}$, a spherical multipole decomposition of the perturbing
field is an efficient way to reduce the number of CSFs to be included.
Consider a spherical multipole operator of angular momentum $J$ and
parity $(-1)^{\Pi}$ ($\Pi=0$ for even and $\Pi=1$ for odd), because
both xenon and germanium initially ($I$) are at $(J_{I},\Pi_{I})=(0,0)$
states, their final ($F$) states will have the same quantum numbers
as the transition operator, i.e., $(J_{F},\Pi_{F})=(J,\Pi)$. Furthermore,
because the transition operator is one-body, the one-particle-one-hole
states it creates ($w_{\alpha+}-u_{\alpha}$) or annihilate ($u_{\alpha}-w_{\alpha-}$)
must couple to the desired state of $(J,\Pi)$.

Let us use the $\hat{C}_{1}$ transition operator $(J,\Pi)=(1,1)$,
which is part of spin-independent DM-electron scattering, as an example.
First, we note that by Wigner-Eckart theorem, all the algebra related
to magnetic quantum numbers can be carried out straightforwardly,
so we abbreviate a quantum label $\alpha$ to its reduced version
$a$. When a $a=5p_{1/2}$ electron is ionized to continuum by $\hat{C}_{1}$,
the possible final partial waves can only be $\epsilon s_{1/2}$ and
$\epsilon d_{3/2}$, with $\epsilon$ fixed by energy conservation.
As a result, the quantum label of $a+$ can be either $\epsilon s_{1/2}$
and $\epsilon d_{3/2}$. On the other hand, for $a=5p_{3/2}$, $a+$
can be $\epsilon s_{1/2}$, $\epsilon d_{3/2}$, or $\epsilon d_{5/2}$.
For other 15 core orbitals of different $a$'s, there are 37 more
distinct $a+$ orbitals which can be mixed. The enumeration for $a-$
states follows the same angular momentum and parity selection rules,
except the energy eigenvalues are different. Altogether, at zeroth
order, the DF orbital equation is a set of 17 coupled integral-differential
equations of $u_{a}(\vec{r})$'s. With the input from zeroth-order
results, the RRPA orbital equation is a set of 84 coupled integral-differential
equations of $w_{a\pm}(\vec{r})$'s and solved by self-consistent
field methods. This is clearly more numerically-intensive task than
the decoupled equations one normally encounter in Hartree frozen core,
or local exchange approximation. On the other hand, as the equation
shows, it is a method that goes beyond the typical mean-field approaches
that consider the correlation effects in both the ground and excited
states. 

Finally, the transition matrix element of a spherical multipole operator
$\hat{O}_{J}^{M}$ is simply calculated by the unperturbed and perturbed
orbitals, $u_{\alpha}$ and $w_{\alpha\pm}$, that make up the initial
and final states, i.e.,

\begin{eqnarray}
\braket{\Psi_{f}|\hat{O}_{J}^{M}|\Psi_{i}} & = & \sum_{\alpha}\Lambda_{\alpha}(\braket{w_{\alpha+}|\hat{O}_{J}^{M}|u_{\alpha}}+\braket{u_{\alpha}|\hat{O}_{J}^{M}|w_{\alpha-}})\,,
\end{eqnarray}
where $\alpha$ run through all the occupied orbitals in the reference
state. For a core orbital, the weighting factor $\Lambda_{\alpha}$
is always 1, and this applies to all xenon orbitals and germanium
orbitals other than $4p$. As we choose the germanium divalent configuration
to be $C_{1}\ket{4p_{1/2}^{2}}+C_{2}\ket{4p_{3/2}^{2}}$, the weighting
factors for $4p_{1/2}$ and $4p_{3/2}$ are $\left|C_{1}\right|^{2}$
and $\left|C_{2}\right|^{2}/2$, respectively, where the extra $1/2$
factor is due to the shell being only half-filled.

\bibliography{MBARF}

\end{document}